\documentclass[aps,pre,twocolumn,superscriptaddress]{revtex4-1}
\usepackage{graphicx}
\usepackage{dcolumn}
\usepackage{bm}
\usepackage{amsmath}
\usepackage{amssymb}

\begin{document}

\title{Binding of anisotropic curvature-inducing proteins onto membrane tubes}

\author{Hiroshi Noguchi}
\email[]{noguchi@issp.u-tokyo.ac.jp}
\affiliation{Institute for Solid State Physics, University of Tokyo, Kashiwa, Chiba 277-8581, Japan}
\author{Caterina Tozzi}
\affiliation{Universitat Polit{\`e}dcnica de Catalunya-BarcelonaTech,
08034 Barcelona, Spain}
\affiliation{Present address: Vall d'Hebron Institute of Oncology (VHIO), 08035 Barcelona, Spain}
\author{Marino Arroyo}
\affiliation{Universitat Polit{\`e}dcnica de Catalunya-BarcelonaTech,
08034 Barcelona, Spain}
\affiliation{Institute for Bioengineering of Catalonia (IBEC), The Barcelona Institute for
Science and Technology (BIST), 08028 Barcelona, Spain}
\affiliation{Centre Internacional de M{\`e}todes Num{\`e}rics en Enginyeria (CIMNE), 08034
Barcelona, Spain}


\begin{abstract}
Bin/Amphiphysin/Rvs superfamily proteins and other curvature-inducing proteins
have anisotropic shapes and  anisotropically bend biomembrane. 
Here, we report how the anisotropic proteins bind the membrane tube and are orientationally ordered
using mean-field theory including an orientation-dependent excluded volume.
The proteins exhibit a second-order or first-order nematic transition
with increasing protein density
depending on the radius of the membrane tube.
The tube curvatures for the maximum protein binding and orientational order
are different and varied by the protein density and rigidity.
As the external force along the tube axis increases,
a first-order transition from a large tube radius with low protein density
to a small radius with high density occurs once,
and subsequently, the protein orientation tilts to the tube-axis direction.
When an isotropic bending energy is used for the proteins with an elliptic shape,
the force-dependence curves become symmetric and the first-order transition occurs twice. 
This theory quantitatively reproduces the results of meshless membrane simulation for short proteins,
whereas deviations are seen for long proteins owing to the formation of protein clusters.
\end{abstract}

\maketitle

\section{Introduction}

In living cells, numerous types of proteins work together to regulate biomembrane shapes of cells and organelles~\cite{mcma05,suet14,joha15,bran13,hurl10,mcma11,baum11,has21}.
Proteins are also involved in dynamic processes such as endo-/exocytosis, vesicle transport, cell locomotion, and cell division.
Clathrin and coat protein complex (COPI and COPII) bend membranes in a laterally isotropic manner and
 generate spherical buds~\cite{joha15,bran13,hurl10,mcma11}.
On the contrary, Bin/Amphiphysin/Rvs (BAR) superfamily proteins bend the membrane anisotropically
and generate cylindrical membrane tubes~\cite{mcma05,suet14,joha15,itoh06,masu10,mim12a,fros08,sorr12,zhu12,tana13,adam15}.
The BAR domains consist of a banana-shaped dimer and bend the membrane along its axis.
Dysfunctions of the BAR proteins are considered to be implicated in neurodegenerative, cardiovascular, and neoplastic diseases.
Thus, understanding the mechanism of the curvature generation by these proteins is important.

The curvature-inducing proteins can sense the membrane curvature, i.e., 
they are concentrated in membranes that have their preferred curvatures.
 The sensing of curvature-inducing proteins, such as BAR proteins~\cite{baum11,has21,sorr12,prev15,tsai21}, dynamin~\cite{roux10}, 
and G-protein coupled receptors~\cite{rosh17}, has been examined using
a tethered vesicle pulled by optical tweezers and a micropipette.
They typically bind more onto the membrane tube than the remaining spherical component.

Theoretically,
the bending energy of a single-component fluid membrane
is well described by the second-order expansion to the curvature (Canham--Helfrich energy)~\cite{canh70,helf73}.
The binding of proteins with a laterally isotropic spontaneous curvature
is considered to locally change the coefficients of the Canham--Helfrich energy (the bending rigidity and spontaneous curvature).
Budding~\cite{lipo92,sens03,fore14,frey20,tozz19,nogu21a} and other shape deformations~\cite{gout21,nogu21b} induced by protein binding 
have been well explained by mean-field theories using this bending energy.
Moreover, traveling waves of membrane deformation can be reproduced by the coupling 
with reaction-diffusion of multiple types of proteins~\cite{gov18,wu18,tame20,tame21}.
In contrast, the effects of the anisotropic spontaneous curvature of proteins have been much less explored.
Instead, the bending energy for isotropic spontaneous curvature has been often used for the analysis of BAR proteins~\cite{prev15,wu18,tsai21}.
A few approaches have been examined for the anisotropy of the protein binding.
The  Canham--Helfrich energy was extended for anisotropic spontaneous curvature \cite{four96,kaba11}
and membrane-mediated interactions between non-deformable anisotropic objects have been investigated~\cite{domm99,domm02,schw15,nogu17}. 
For cylindrical membranes, the axis of banana-shaped proteins were assumed aligned in the azimuthal direction
to derive a force--extension curve~\cite{nogu15b}. 
However, the entropic interaction of the protein orientation has not been considered in these studies.

Recently, this entropic interaction was taken into account by two of us and co-workers~\cite{tozz21,roux21} based on Nascimentos' theory for three-dimensional liquid-crystals~\cite{nasc17}.
An isotropic-to-nematic transition was obtained on a fixed membrane shape.
In this study, we examine the binding of the anisotropic proteins to a cylindrical membrane tube in detail.
The axial force along the membrane tube and equilibrium of protein binding/unbinding are  considered.
Moreover, we clearly show the difference from the binding of isotropic proteins~\cite{nogu21b}. 
The tube part of a tethered vesicle is well approximated by this tube with no volume constraint,
 when tube radius is much smaller than the vesicle radius~\cite{nogu21b}.

Several types of membrane models have been developed for coarse-grained simulations~\cite{muel06,vent06,nogu09}.
The protein binding has been investigated using molecular simulations~\cite{arkh08,yu13,simu13,gome16,olin16,take17,mahm19}, 
dynamically triangulated membrane simulations~\cite{rama13,behe21}, 
and meshless membrane simulations~\cite{nogu14,nogu15b,nogu16,nogu16a,nogu17,nogu17a,nogu19a}.
Among them, however, the binding effects on the axial force of a membrane tube have been investigated only by the meshless simulations~\cite{nogu14,nogu15b,nogu16a};
a characteristic force dependence on the protein curvature was reported for homogeneous states at low protein curvatures,
 in addition to the protein assembly accompanied by membrane shape transformation at high protein curvatures.
Here, we compare our theoretical results with those of the meshless simulations.

The mean-field theory is described in Sec.~\ref{sec:theory}.
Simulations of membrane tubes are described in Sec.~\ref{sec:sim}.
The simulation results are compared with the theoretical results in Sec.~\ref{sec:sres}.
Finally, a summary and discussion are presented in Sec.~\ref{sec:sum}.

\begin{figure}
\includegraphics[]{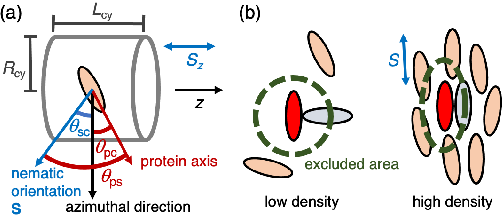}
\caption{
Schematic of the theoretical model.
(a) An elliptic protein on a membrane tube.
The angles between the nematic direction {\bf S}, the azimuthal direction, and/or
 and protein axis.
(b) Excluded-volume interactions between proteins.
A perpendicular protein pair has a larger excluded area (represented by thick dashed lines)
than a parallel pair, leading to a nematic order at a high density.
}
\label{fig:cat}
\end{figure}

\section{Theoretical analysis}\label{sec:theory}
\subsection{Theory}\label{sec:freeen}

Protein binding on 
a cylindrical membrane tube is considered as depicted in Fig.~\ref{fig:cat}(a).
The membrane is in a fluid phase and the surface area $A$ is fixed.
The radius and length of the tube are $R_{\rm cy}$ and $L_{\rm cy}$: $A= 2\pi R_{\rm cy}L_{\rm cy}$.
The tube volume can be freely changed. This corresponds to the tubular region of a tethered vesicle in the limit condition,
in which the tube volume is negligibly small ($ \pi R_{\rm cy}^2L_{\rm cy} \ll V$, where $V$ is the vesicle volume)~\cite{smit04,nogu21b}.
Proteins can align in the membrane surface.
To quantify it, the degree of the orientational order is calculated as
\begin{eqnarray} \label{eq:s}
S &=& 2 \langle s_{\rm p}(\theta_{\rm ps}) \rangle, \\
s_{\rm p}(\theta_{\rm ps}) &=& \cos^2(\theta_{\rm ps}) - \frac{1}{2},
\end{eqnarray} 
where $\langle ... \rangle$ is the ensemble average,
and $\theta_{\rm ps}$ is the angle between the major protein axis and the ordered direction.
The angles between  the ordered direction and the azimuthal direction of the cylinder
and between the major protein axis and azimuthal direction are $\theta_{\rm sc}$ and  $\theta_{\rm pc}$, respectively,
with $\theta_{\rm ps}=\theta_{\rm pc}+\theta_{\rm sc}$
 (see Fig.~\ref{fig:cat}(a)).
Experimentally, the oritational order $S_z$ along the tube ($z$) axis is more easily measured:
for $\theta_{\rm sc}=0$ and $\pi/2$, $S_z= - S$ and $S_z= S$, respectively.

The bound protein is approximated to have laterally an elliptic shape 
with an aspect ratio of $d_{\rm el}=\ell_1/\ell_2$ and area $a_{\rm p} = \pi \ell_1\ell_2/4$,
where $\ell_1$ are $\ell_2$ are the lengths in the major and minor axes, respectively.
An orientation-dependent excluded-volume interaction is considered between proteins.
When two proteins are perpendicularly oriented,
the excluded area $A_{\rm exc}$ between them is larger than the parallel pairs, as shown in Fig.~\ref{fig:cat}(b).
This area  $A_{\rm exc}$ is expressed as $A_{\rm exc}= B_0 + B_2(\cos^2(\theta_{\rm pp})-1/2) + O(\cos^4(\theta_{\rm pp}))$ by a Taylor expansion,
where $\theta_{\rm pp}$ is the angle between the major axes of two ellipses.
In our previous study~\cite{tozz21}, the values of $B_0$ and $B_2$ are calculated by a two-parameter fit.
In this study, the one-parameter fit of $A_{\rm exc}= [4 - b_{\rm exc}(\cos^2(\theta_{\rm pp})-1)]a_{\rm p}$ is used,
since the minimum value $A_{\rm exc}^{\rm min}=4a_{\rm p}$ is obtained at the parallel pairs ($\theta_{\rm pp} = 0$) for any ratio of $d_{\rm el}$:
$b_{\rm exc}= 0.840$, $1.98$, $3.44$, and $7.61$ at $d_{\rm el}=2$, $3$, $4$, and $7$, respectively.
The effective excluded area is represented by  $A_{\rm eff}= \lambda A_{\rm exc}$.
The parameter $\lambda$ is a function of the protein density and
 $\lambda=1/2$ at a low-density limit~\cite{nasc17}.
At the close-packed condition, the area fraction $\phi$ of the bound protein has the maximum: 
$\phi_{\rm max}=a_{\rm p}/\lambda A_{\rm exc}^{\rm min} = 1/4\lambda= \pi/2\sqrt{3}\approx 0.907$ in two-dimensional space~\cite{tane97}.
For simplicity, we use a constant value, $\lambda = 1/3$, as in our previous study~\cite{tozz21,roux21}, i.e., $\phi_{\rm max}=0.75$.
In this study, we consider no attractive interactions between the proteins and 
focus on isotropic and nematic phases, such that smectic and crystal phases are not in the scope.

The bending energy of the bare (unbound) membrane is given by
\begin{equation}
U_{\rm mb} = \int  \frac{\kappa_{\rm d}}{2}(C_1+C_2)^2 dA   = \frac{\kappa_{\rm d}A}{2R_{\rm cy}^2},
\end{equation}
where $C_1$ and $C_2$ are the principal curvatures ($C_1=1/R_{\rm cy}$ and $C_2=0$ for the cylinder).
The unbound membrane has a bending rigidity $\kappa_{\rm d}$ and zero spontaneous curvature.
The tubular membrane is connected to a large lipid reservoir, and the area difference elasticity~\cite{seif97,svet09} is negligible.
The bound protein gives an additional bending energy as $\langle U \rangle= U_{\rm mb} + N_{\rm p} \langle U_{\rm p}\rangle$,
where $N_{\rm p}= \phi A/a_{\rm p}$ is the number of the bound protein and $U_{\rm p}$ is the bending energy of one protein.
The protein has an anisotropic bending energy:
\begin{eqnarray}
U_{\rm p} &=&   \frac{\kappa_{\rm p}a_{\rm p}}{2}(C_{\ell 1} - C_{\rm p})^2 + \frac{\kappa_{\rm side}a_{\rm p}}{2}(C_{\ell 2} - C_{\rm side})^2, \\
C_{\ell 1} &=& C_1 \cos^2( \theta_{\rm pc} ) + C_2  \sin^2( \theta_{\rm pc}), \\
C_{\ell 2} &=& C_1 \sin^2( \theta_{\rm pc} ) + C_2  \cos^2(\theta_{\rm pc}),
\end{eqnarray}
where $C_{\ell 1}$ and $C_{\ell 2}$  are curvatures along the major and minor axes of the proteins, respectively.
$\kappa_{\rm p}$ and $C_{\rm p}$ are the bending rigidity and spontaneous curvature along the major protein axis, respectively,
and  $\kappa_{\rm side}$ and $C_{\rm side}$ are those along the minor axis (side direction).
Here, $\kappa_{\rm side}=0$ is used unless  otherwise specified.

The free energy $F_{\rm p}$ of the bound proteins is expressed as
\begin{eqnarray}
F_{\rm p} &=& \int f_{\rm p}\ dA, \\ 
f_{\rm p} &=&  \frac{\phi k_{\rm B}T}{a_{\rm p}}\Big[\ln(\phi) + \frac{S \Psi}{2} - \ln\Big(\int_{-\pi/2}^{\pi/2} w(\theta_{\rm ps})\ d\theta_{\rm ps}\Big)\Big], \ \ \\
w(\theta_{\rm ps})  &=&  g\exp[\Psi s_{\rm p}(\theta_{\rm ps}) + \bar{\Psi}\sin(\theta_{\rm ps})\cos(\theta_{\rm ps}) - \beta U_{\rm p} ]\Theta(g), \\
g   &=& 1-\phi (b_0-b_2S s_{\rm p}(\theta_{\rm ps})),
\end{eqnarray}
where 
$\Theta(x)$ denotes the unit step function, $k_{\rm B}T$ is the thermal energy, and $\beta=1/k_{\rm B}T$.
The factor $g$ expresses the effect of the orientation-dependent excluded volume, where 
 $b_0=B_0\lambda/a_{\rm p}= (4 + b_{\rm exc}/2)\lambda$ and $b_2= -B_2\lambda/a_{\rm p}= b_{\rm exc}\lambda$.
Unoverlapped states exist at $g>0$.
Since $w(\theta_{\rm ps})$ is the weight of each protein orientation,
the ensemble average of a quantity $\chi$ is given by
\begin{eqnarray} \label{eq:av}
\langle \chi \rangle = \frac{\int_{-\pi/2}^{\pi/2} \chi w(\theta_{\rm ps})\ d\theta_{\rm ps} }{\int_{-\pi/2}^{\pi/2}  w(\theta_{\rm ps}) \ d\theta_{\rm ps}}.
\end{eqnarray}
The quantities $\Psi$ and $\bar{\Psi}$ are the symmetric  and asymmetric components of the nematic tensor, respectively, 
and are determined by Eq.~(\ref{eq:s}) and $\langle \sin(\theta_{\rm ps})\cos(\theta_{\rm ps}) \rangle =0$ using Eq.~(\ref{eq:av}).
When the nematic order is parallel to one of the directions of the membrane principal curvatures ($\theta_{\rm sc}=0$ or $\pi/2$),
$\bar{\Psi}=0$.
The free energy minimum is calculated from $\partial f_{\rm p}/\partial S=\partial f_{\rm p}/\partial \theta_{\rm sc}=0$.
More detail is described in Ref.~\citenum{tozz21}.

In this study, we examine the axial force  $f_{\rm ex}$ and the equilibrium of the protein binding and unbinding.
In experiments,
an external force $f_{\rm ex}$ is imposed by optical tweezers and micropipette
in order to extend a membrane tube from a liposome.
The free energy is give by $F=F_{\rm p} + U_{\rm mb} - f_{\rm ex}L_{\rm cy}$.
This force $f_{\rm ex}$ is balanced with the membrane axial force and
 is obtained by $\partial F/\partial L_{\rm cy} =0$ as
\begin{eqnarray}
\label{eq:fex}
f_{\rm ex} = 2\pi\frac{\partial f_{\rm p}}{\partial (1/R_{\rm cy})} + f_{\rm mb}.
\end{eqnarray}
The last term $f_{\rm mb}$ represents the force of the bare membrane tube ($\phi=0$),
\begin{equation} \label{eq:fmb}
f_{\rm mb}=\frac{2\pi\kappa_{\rm d}}{R_{\rm cy}} = \frac{f_0}{R_{\rm cy}C_{\rm p}}, 
\end{equation}
where $f_0=2\pi\kappa_{\rm d}C_{\rm p}$ is the force at $R_{\rm cy}C_{\rm p}=1$ 
and is used as the unit hereafter.

The proteins bind  and unbind the membrane with the binding chemical potential $\mu$.
The equilibrium of the binding and unbinding is obtained by
minimizing $F - \mu N_{\rm p}$.
Hence, 
the equilibrium protein density is calculated from 
$\mu = a_{\rm p}\partial f_{\rm p}/\partial \phi$.
Here, the number $N_{\rm lip}$ of the lipids and the area $A$ remain constant,
so that the ensemble is changed from the $N_{\rm p} N_{\rm lip}AT$ ensemble to $\mu N_{\rm lip}AT$ ensemble.

Unless otherwise specified,
we use $d_{\rm el}=3$ and $a_{\rm p}C_{\rm p}^2=0.26$,
which correspond to the N-BAR domain ($\ell_1=15$ nm $\ell_2=5$ nm, and $1/C_{\rm p}=15$ nm)~\cite{roux21}.
Another area ratio $a_{\rm p}C_{\rm cy}^2=0.26$ is  used to examine $C_{\rm p}$ dependence.
The detail of the numerical methods is described in Appendix~\ref{sec:psi}.

\begin{figure}
\includegraphics[]{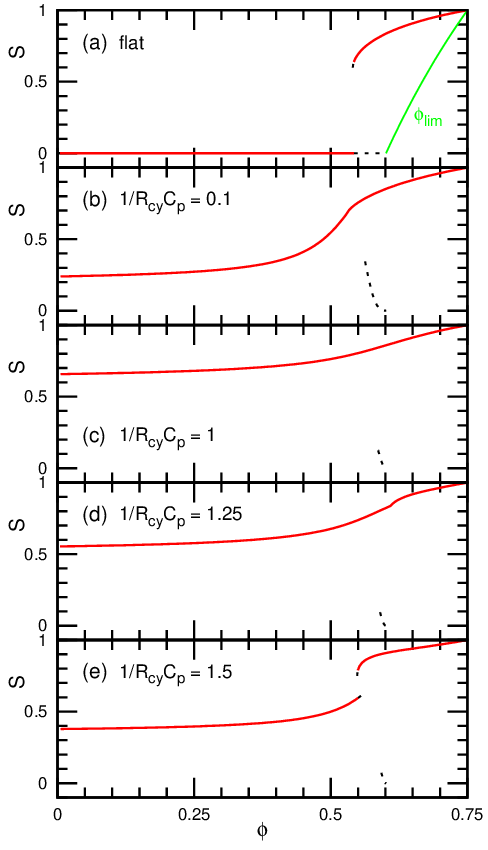}
\caption{
Orientational degree $S$ of the proteins for $1/R_{\rm cy}C_{\rm p}=0$ (flat membrane), 
$0.1$, $1$, $1.25$, and $1.5$  at $\kappa_{\rm p}/k_{\rm B}T=40$ and  $d_{\rm el}=3$.
The solid and dashed lines represent the data of stable and metastable states, respectively.
The right line in (a) represents the maximum density $\phi_{\rm lim}(S)$  given by Eq.~(\ref{eq:pmax}).
}
\label{fig:cy}
\end{figure}

\begin{figure}
\includegraphics[]{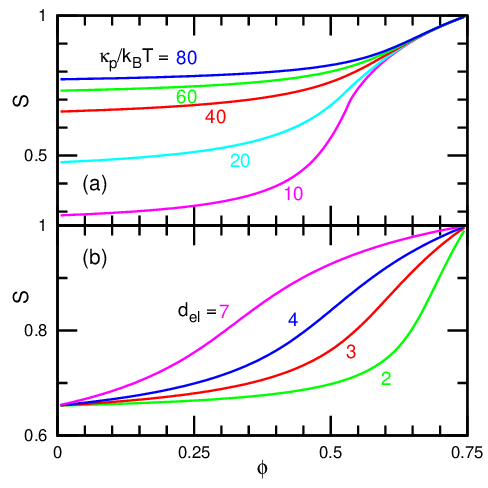}
\caption{
Effects of (a) the bending rigidity $\kappa_{\rm p}$ and (b) elliptic ratio $d_{\rm el}$ of the proteins
on the density $\phi$ dependence of the orientational degree $S$ at $1/R_{\rm cy}C_{\rm p}=1$.
(a) From top to bottom, $\kappa_{\rm p}/k_{\rm B}T=80$, $60$, $40$, $20$, and $10$ at $d_{\rm el}=3$.
(b) From top to bottom, $d_{\rm el}=7$,  $4$, $3$, and $2$ at  $\kappa_{\rm p}/k_{\rm B}T=40$.
The metastable states at $\phi\simeq \phi_{\rm lim}(0)$ are not shown.
}
\label{fig:kp}
\end{figure}

\begin{figure}
\includegraphics[]{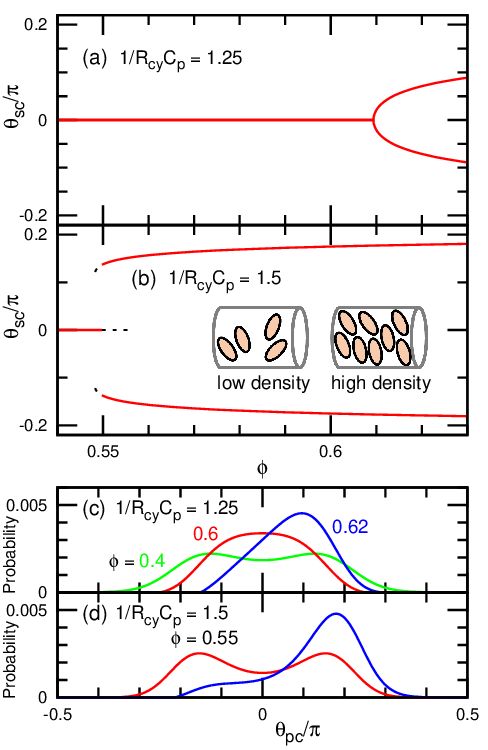}
\caption{ 
Angles $\theta_{\rm sc}$ and $\theta_{\rm pc}$ for $1/R_{\rm cy}C_{\rm p}=1.25$ and $1.5$  at $\kappa_{\rm p}/k_{\rm B}T=40$ and  $d_{\rm el}=3$.
Second order and first order transitions occur for $1/R_{\rm cy}C_{\rm p}=1.25$ and $1.5$, respectively.
The solid lines in (a),(b) and dashed lines in (b) represent the data of stable and metastable states, respectively.
(d) Two states coexist at $\phi=0.55$ and $1/R_{\rm cy}C_{\rm p}=1.5$. 
In the inset of (b), the protein states are schematically depicted for low and high protein densities $\phi$.
}
\label{fig:ang}
\end{figure}

\begin{figure}
\includegraphics[]{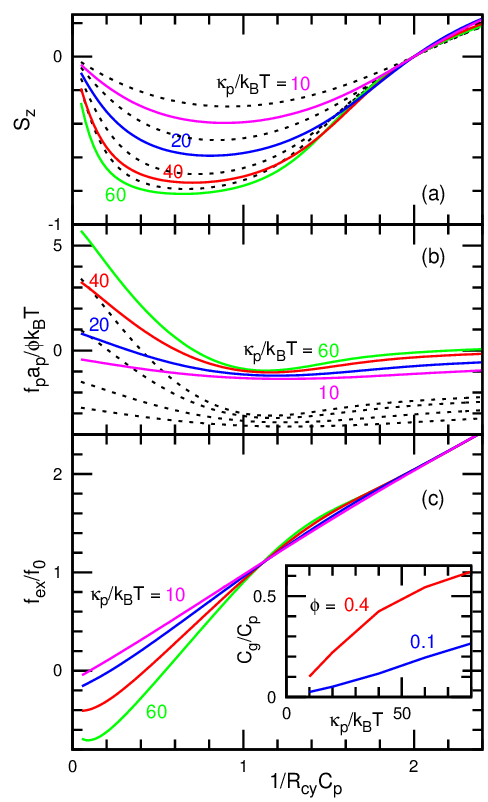}
\caption{
Dependence on the radius $R_{\rm cy}$ of the membrane tube for a constant protein density
at $\kappa_{\rm p}/k_{\rm B}T=40$ and $d_{\rm el}=3$.
(a) Orientational degree $S_z$ along the tube axis.
(b) Free energy density $f_{\rm p}$ of the proteins.
(c) Axial force $f_{\rm ex}$ normalized by $f_0=2\pi \kappa_{\rm d} C_{\rm p}$.
The solid and dashed lines represent the data at $\phi=0.4$ and $0.1$, respectively.
(a),(c) From top to bottom, $\kappa_{\rm p}/k_{\rm B}T=10$, $20$, $40$, and $60$.
(b) From top to bottom, $\kappa_{\rm p}/k_{\rm B}T=60$, $40$, $20$, and $10$.
The inset in (c) shows the generation curvature $C_{\rm g}$ (tube curvature at $f_{\rm ex}=0$) 
for $\phi=0.1$ and $0.4$.
}
\label{fig:ds}
\end{figure}

\begin{figure}
\includegraphics[]{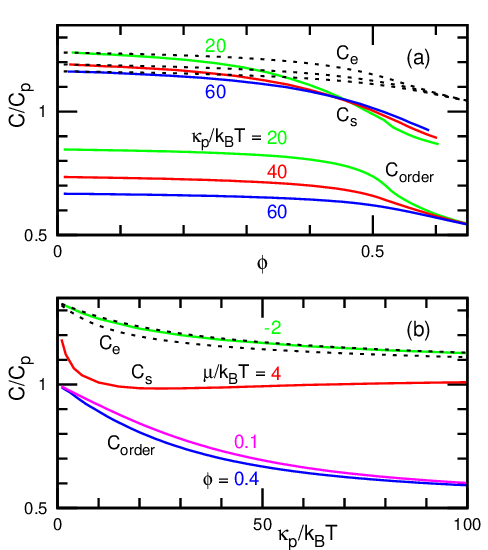}
\caption{
Order curvature $C_{\rm order}$ (maximum of $S$), free-energy-minimum curvature $C_{\rm e}$, and 
sensing curvature $C_{\rm s}$ (maximum binding) of the proteins at $d_{\rm el}=3$.
(a) Dependence on the density $\phi$. From top to bottom, $\kappa_{\rm p}/k_{\rm B}T=20$, $40$, and $60$.
(b) Dependence on the protein rigidity $\kappa_{\rm p}$.
From top to bottom, $\phi=0.1$ and $0.4$ for $C_{\rm order}$ and $C_{\rm e}$,
and $\mu/k_{\rm B}T=-2$ and $4$ for $C_{\rm s}$.
The dashed lines represent  $C_{\rm e}$.
}
\label{fig:s0}
\end{figure}

\begin{figure}
\includegraphics[]{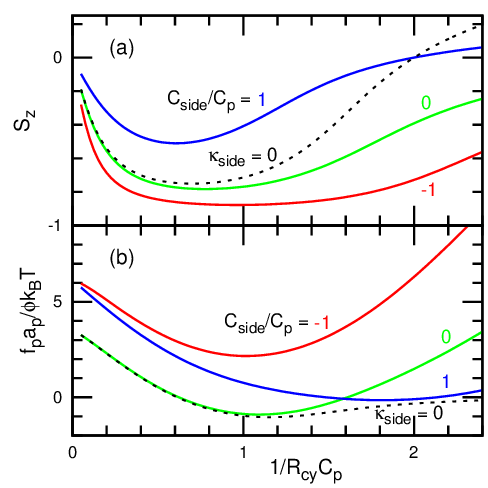}
\caption{
Effects of the spontaneous curvature $C_{\rm side}$ in the side direction of the proteins
on the $R_{\rm cy}$ dependence of (a) the orientational degree $S_z$ along the tube axis and (b) free energy density $f_{\rm p}$ of the proteins
at $\phi=0.4$, $\kappa_{\rm p}/k_{\rm B}T=40$, and $d_{\rm el}=3$.
The solid lines represent the data for  $C_{\rm side}/C_{\rm p}= -1$, $0$, and $1$ at $\kappa_{\rm side}/k_{\rm B}T=20$.
The dashed lines represent the data for no side bending energy ($\kappa_{\rm side}=0$).
}
\label{fig:side}
\end{figure}

\begin{figure}
\includegraphics[]{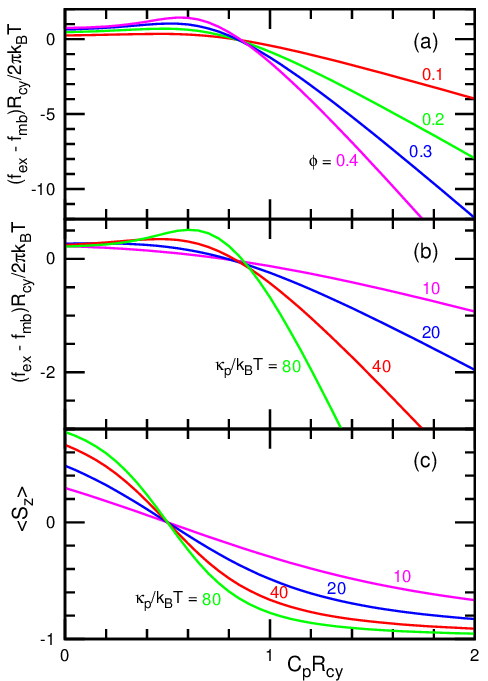}
\caption{
Dependence on the protein curvature $C_{\rm p}$ at $a_{\rm p}C_{\rm cy}^2=0.26$ and $d_{\rm el}=3$.
The force  $f_{\rm ex}$ for four values of the protein density $\phi$ and rigidity $\kappa_{\rm p}$ are shown 
in (a) and (b), respectively.
(c) Orientational degree $S_z$ for the same data in (b).
(a) $\phi=0.1$, $0.2$, $0.3$, and $0.4$ at  $\kappa_{\rm p}/k_{\rm B}T=40$.
(b),(c) $\kappa_{\rm p}/k_{\rm B}T=10$, $20$, $40$, and $80$ at $\phi=0.1$.
}
\label{fig:fcp}
\end{figure}

\begin{figure}
\includegraphics[]{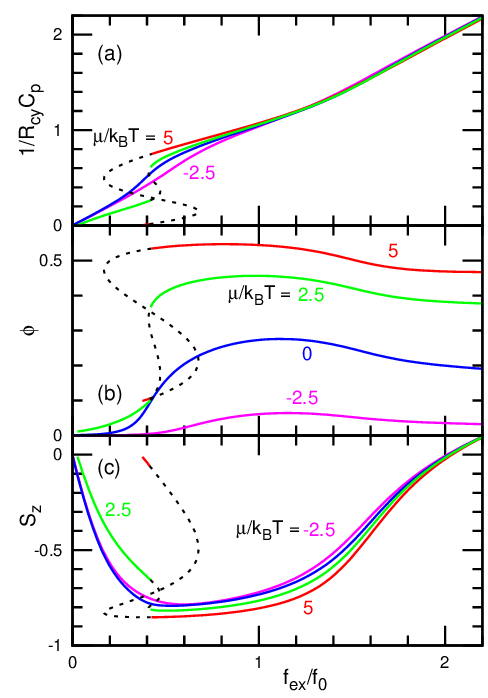}
\caption{
Force $f_{\rm ex}$ dependence of (a) the curvature $1/R_{\rm cy}$ of the cylindrical membrane, 
(b) protein density $\phi$, and  (c) the orientational degree $S_z$ along the tube axis
for $\mu/k_{\rm B}T=-2.5$, $0$, $2.5$, and $5$ at $d_{\rm el}=3$ and  $\kappa_{\rm p}/k_{\rm B}T=60$.
The solid lines represent thermal equilibrium states.
The dashed lines represent the metastable and free-energy-barrier states.
The force is normalized by $f_0=2\pi \kappa_{\rm d} C_{\rm p}$.
}
\label{fig:muk60}
\end{figure}

\begin{figure}
\includegraphics[]{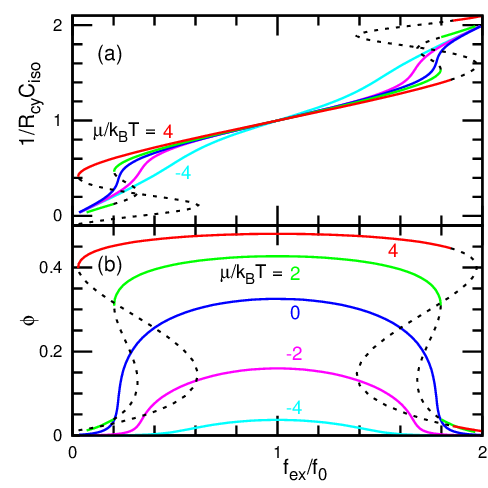}
\caption{
Force $f_{\rm ex}$ dependence for the proteins with the isotropic bending energy $U_{\rm iso}$
with $\mu/k_{\rm B}T=-4$, $-2$, $0$, $2$, and $4$ at $d_{\rm el}=3$ and $\kappa_{\rm iso}/k_{\rm B}T=60$.
(a) Curvature $1/R_{\rm cy}$ of the membrane tube.
(b) Protein density $\phi$.
The solid lines represent thermal equilibrium states.
The dashed lines represent the metastable and free-energy-barrier states.
The force is normalized by $f_0=2\pi \kappa_{\rm d} C_{\rm iso}$.
}
\label{fig:muiso}
\end{figure}

\subsection{Theoretical results}\label{sec:tres}

For flat membranes,
the proteins exhibit a first-order transition 
from a randomly oriented state ($S=0$) to
an ordered state ($S>0$) with increasing protein density $\phi$, as shown in Fig.~\ref{fig:cy}(a).
This transition density decreases as the elliptic ratio $d_{\rm el}$ increases,
and the same behavior is obtained for spherical membranes~\cite{tozz21}.
For cylindrical membranes,
the proteins are oriented on average even at $\phi \to 0$ (see Figs.~\ref{fig:cy}(b)--(e)).
The maximum density $\phi_{\rm lim}(S)$ is given by 
\begin{equation}
\label{eq:pmax}
\phi_{\rm lim}(S) = \frac{1}{b_0- b_2 S/2},
\end{equation}
that is independent of the membrane curvature ($\phi_{\rm max}=\phi_{\rm lim}(1)$).
The straight line ($S=0$) for $0<\phi<\phi_{\rm lim}(0)$ in  Fig.~\ref{fig:cy}(a)
is divided into two, and the right branch remains even at large tube curvatures, $1/R_{\rm cy}$,
although it has a high energy with a narrow width of $\phi$ (see dashed lines at $\phi\simeq 0.6$ in Figs.~\ref{fig:cy}(b)--(e)).
Meanwhile, the left branch connects to the upper branch at $1/R_{\rm cy}C_{\rm p} \gtrsim 0.1$ (see Fig.~\ref{fig:cy}(b)).
Note that it is separated at $1/R_{\rm cy}C_{\rm p} = 0.01$ (data not shown).
At $1/R_{\rm cy}C_{\rm p} \leq 1$, the proteins prefer to align to the azimuthal direction ($\theta_{\rm pc}=0$), 
while the thermal fluctuations disturb it.
Hence, the proteins are more ordered (higher $S$) at higher $\kappa_{\rm p}$ (see Fig.~\ref{fig:kp}(a)).
At high density $\phi$ (close to $\phi_{\rm max}$), 
$S$ is dominantly determined by the orientation-dependent excluded volume
and the effects increase with increasing $d_{\rm el}$ (see Fig.~\ref{fig:kp}).

At $1/R_{\rm cy}C_{\rm p} > 1$, the protein preferred direction is tilted 
either to a positive or negative angle of $\theta_{\rm pc}= \pm \arccos(\sqrt{R_{\rm cy}C_{\rm p}})$.
At low $\phi$, the positive and negative angles simultaneously exist, 
so that the proteins exhibit a symmetric distribution with $\theta_{\rm sc}=0$ (see Fig.~\ref{fig:ang}).
In contrast, at high $\phi$, these two angles cannot coexist at the same time 
owing to the large excluded-volume interactions between them
(see the right distributions in Figs.~\ref{fig:ang}(c) and (d)).
The transitions between these two states  are the second order and first order 
for $1/R_{\rm cy}C_{\rm p} \leq 1.3$ and $1/R_{\rm cy}C_{\rm p} \geq 1.35$ (see Figs.~\ref{fig:ang}(a) and (b)), respectively. 
The two states coexist at $\phi\simeq 0.55$ and $1/R_{\rm cy}C_{\rm p} = 1.5$ as shown in Fig.~\ref{fig:ang}(d).
Correspondingly, 
the $S$--$\phi$ curves exhibit discrete changes of the slope and position
(see Figs.~\ref{fig:cy}(d) and (e)), respectively.
In the case of the second-order transition, the excluded-volume interactions push the protein 
into the azimuthal direction leading to the angular distribution of a single peak 
near the transition point (see the data at $\phi=0.6$ in Fig.~\ref{fig:ang}(c)),
so that the symmetric peak continuously changes to an asymmetric peak at the transition.

Figure~\ref{fig:ds} shows the tube curvature $1/R_{\rm cy}$ dependence of 
the orientational degree $S_z$ along the tube axis, protein free-energy, and axial force $f_{\rm ex}$.
The preferred orientation changes from $\theta_{\rm sc}=0$ to $\pi/2$ at $1/R_{\rm cy}C_{\rm p}=2$,
so that  $S_z$ changes from negative values ($S_z=-S$) to positive values ($S_z=S$).
Interestingly, the orientational order ($S=|S_z|$) has a maximum at $1/2<1/R_{\rm cy}C_{\rm p}<1$,
i.e., less than the matching curvature $1/R_{\rm cy}C_{\rm p}=1$ (see Figs.~\ref{fig:ds}(a) and \ref{fig:cy}).
Oppositely, the curvature $C_{\rm e}$ of the free-energy minimum is 
higher than the matching curvature (see Fig.~\ref{fig:ds}(b)).
These are determined by the competition between the orientational entropy and bending energy.
The curvature ($C_{\rm order}$) of the maximum order and $C_{\rm e}$ decrease 
with increasing $\phi$ and $\kappa_{\rm p}$ (see Fig.~\ref{fig:s0});
at $\kappa_{\rm p}\to \infty$, $C_{\rm order}\to 1/2R_{\rm cy}$, in which the strength of the approximately harmonic potential of $U_{\rm p}$ for $\theta_{\rm pc}\ll 1$ (i.e., $\partial^2 U_{\rm p}/\partial \theta_{\rm pc}^2|_{\theta_{\rm pc}=0}$) is maximum.
The amplitudes of the order and the depth of free-energy minimum increase with increasing $\kappa_{\rm p}$ 
(see Figs.~\ref{fig:ds}(a) and (b), respectively).
Corresponding to the larger energy change in $f_{\rm p}$,
the axial force $f_{\rm ex}$ deviates more from the value
of the bare membrane ($f_{\rm mb}$ given by Eq.~(\ref{eq:fmb})) at higher $\kappa_{\rm p}$ (see Fig.~\ref{fig:ds}(c)).
Spontaneously formed membrane tubes require no axial force (i.e., $f_{\rm ex}=0$).
The generation curvature $C_{\rm g}$ of this spontaneous tube increases 
with increasing $\kappa_{\rm p}$ and $\phi$ (see the inset of Fig.~\ref{fig:ds}(c)),
i.e., a narrower tube is generated.
Note that $C_{\rm g}$ also depends on the bending rigidity $\kappa_{\rm d}$ of the bare membrane 
in contrast to $C_{\rm order}$ and $C_{\rm e}$.

Next, we examine the effects of the protein bending energy in the side direction
(see Fig.~\ref{fig:side}).
At zero side spontaneous curvature ($C_{\rm side}=0$),
the proteins more align in the azimuthal direction with increasing $\kappa_{\rm side}$,
since the curvature $C_{\ell 2}$ in the side direction becomes closer to $C_{\rm side}$.
This effect is pronounced at narrow tubes, whereas it is negligible for $1/R_{\rm cy}C_{\rm p}<1$.
For a negative value of $C_{\rm side}$,
$C_{\rm order}$ and $C_{\rm e}$ become close to $C_{\rm p}$,
and the proteins more align in a wider range of $1/R_{\rm cy}$.
For a positive value of $C_{\rm side}$,
$C_{\rm order}$ and $C_{\rm e}$ become deviated from $C_{\rm p}$,
and the proteins less align.
The generation curvature slightly increases with increasing $C_{\rm side}$:
$C_{\rm g}= 0.412$, $0.421$, and $0.448$ at $C_{\rm side}=-1$, $0$, and $1$, respectively, 
for the condition used in Fig.~\ref{fig:side}.

We have fixed the protein curvature $C_{\rm p}$ until here.
Figure~\ref{fig:fcp} shows the effects of $C_{\rm p}$ variation with maintaining the other parameters.
As $C_{\rm p}$ changes from null to $1/R_{\rm cy}$,
the nematic orientation changes from the axial direction ($\theta_{\rm sc}=\pi/2$) to the azimuthal direction ($\theta_{\rm sc}=0$)
(see the left half of Fig.~\ref{fig:fcp}(c)).
This change becomes steeper at higher $\kappa_{\rm p}$.
During this change, the axial force $f_{\rm ex}$ is almost constant, although a small peak appears for high $\kappa_{\rm p}$ and/or high $\phi$ (see  the left regions of Figs.~\ref{fig:fcp}(a) and (b)).
This is due to little change in the bending energy, 
because the proteins can find their preferred curvature by adjusting their orientation.
For $C_{\rm p} \gtrsim 1/R_{\rm cy}$,  $f_{\rm ex}$ almost linearly decreases,
and the slope increases with increasing $\kappa_{\rm p}$ and $\phi$ (see the right region of Figs.~\ref{fig:fcp}(a) and (b)).
These dependencies qualitatively agree with the results of our previous meshless membrane simulations~\cite{nogu14,nogu15b,nogu16a}.
A quantitative comparison is described in Sec.~\ref{sec:sres}.

Finally, we examine the equilibrium of the protein binding and unbinding.
As the binding chemical potential $\mu$ increases,
more proteins bind onto the membrane.
The protein binding exhibits a first-order transition 
from a wide tube with low $\phi$ to a narrow tube with high $\phi$ at small force, $f_{\rm ex}<f_0$ 
(see Fig.~\ref{fig:muk60}).
This transition agrees with the observation of the coexistence of two tubes with different $R_{\rm cy}$ and $\phi$ in the experiments of an I-BAR protein~\cite{prev15}.
The force-dependence curves shown in Fig.~\ref{fig:muk60} are asymmetric
and exhibit weak dependence at $f_{\rm ex}>f_0$,
owing to
the adjustment of the protein orientation that reduces a change in the protein bending energy (see Fig.~\ref{fig:muk60}(c)).
These behaviors are different from the binding of proteins with an isotropic spontaneous curvature~\cite{nogu21b},
where the $f_{\rm ex}$--$1/R_{\rm cy}$ and$f_{\rm ex}$--$\phi$ curves are point symmetric and reflection symmetric
to $f_{\rm ex}=f_0$, respectively.

The protein binding has a maximum in the variation of the tube curvature (compare Figs.~\ref{fig:muk60}(a) and (b)).
This curvature is called sensing curvature (denoted $C_{\rm s}$) 
and can  be calculated from $\partial\phi/\partial (1/R_{\rm cy})=0$.
Interestingly, $C_{\rm s}$ is varied by $\mu$ and $\kappa_{\rm p}$ (see Fig.~\ref{fig:s0}).
For low $\phi$ at low $\mu$, $C_{\rm s}$ approach $C_{\rm e}$, since the excluded volume gives negligible effects.
For high $\mu$ ($\phi \gtrsim 0.5$ in Fig.~\ref{fig:s0}(a)),  $C_{\rm s}$ becomes lower than $C_{\rm p}$, and $\phi$ has a broad peak.
A similar $C_{\rm s}$ dependence on the tube curvature has been reported in the experiments of the BAR proteins~\cite{prev15,tsai21}.
It indicates the anisotropic interaction of the BAR proteins.
 
The asymmetry of the force-dependence curves is caused not by the orientation-dependent excluded volume 
but by the anisotropy of the protein bending energy.
To clearly show it,
the force-dependence curves for the elliptic proteins with an isotropic spontaneous curvature $C_{\rm iso}$
are plotted in Fig.~\ref{fig:muiso}.
The proteins have a bending energy 
\begin{equation}
U_{\rm iso} =   \frac{\kappa_{\rm iso}a_{\rm p}}{2}(C_1 + C_2 - C_{\rm iso})^2,
\end{equation}
instead of $U_{\rm p}$.
Note that the anisotropic bending energy $U_{\rm p}$ with $\kappa_{\rm p}=\kappa_{\rm side}$ and $C_{\rm p}=C_{\rm side}$ 
does not coincide to $U_{\rm iso}$ except for the case of $\theta_{\rm pc}=0$ or $\pi/2$.
The $f_{\rm ex}$--$1/R_{\rm cy}$ and$f_{\rm ex}$--$\phi$ curves become point symmetric and reflection symmetric
to $f_{\rm ex}=f_0$, respectively,
and the first-order transitions occur both at small and large forces symmetrically.
The transition points are almost constant for a variation in $\mu$.
This is due to the  excluded-volume dependence on the protein orientation,
since the transition points move outwards in the case of orientation-independent excluded volume~\cite{nogu21b}.

\section{Simulation}\label{sec:sim}

\subsection{Simulation model}

A fluid membrane is represented by a self-assembled single-layer sheet of $N$ particles.
The position and orientational vectors of the $i$-th particle are ${\bm{r}}_{i}$ and ${\bm{u}}_i$, respectively.
The membrane particles interact with each other via a potential $U=U_{\rm {rep}}+U_{\rm {att}}+U_{\rm {bend}}+U_{\rm {tilt}}$.
The potential $U_{\rm {rep}}$ is an excluded volume interaction with diameter $\sigma$ for all pairs of particles.
The solvent is implicitly accounted for by an effective attractive potential $U_{\rm {att}}$. 
The details of the meshless membrane model and protein rods are described 
in Ref.~\citenum{shib11} and Refs.~\citenum{nogu14,nogu16a}, respectively.
We employ the parameter sets used in Ref.~\citenum{nogu16a}.

The bending and tilt potentials
are given by
 $U_{\rm {bend}}/k_{\rm B}T=(k_{\rm {bend}}/2) \sum_{i<j} ({\bm{u}}_{i} - {\bm{u}}_{j} - C_{\rm {bd}} \hat{\bm{r}}_{i,j} )^2 w_{\rm {cv}}(r_{i,j})$
and $U_{\rm {tilt}}/k_{\rm B}T=(k_{\rm{tilt}}/2) \sum_{i<j} [ ( {\bm{u}}_{i}\cdot \hat{\bm{r}}_{i,j})^2 + ({\bm{u}}_{j}\cdot \hat{\bm{r}}_{i,j})^2  ] w_{\rm {cv}}(r_{i,j})$, respectively,
where ${\bm{r}}_{i,j}={\bm{r}}_{i}-{\bm{r}}_j$, $r_{i,j}=|{\bm{r}}_{i,j}|$,
 $\hat{\bm{r}}_{i,j}={\bm{r}}_{i,j}/r_{i,j}$, $w_{\rm {cv}}(r_{i,j})$ is a weight function.
The spontaneous curvature $C_0$ of the membrane is 
given by $C_0\sigma= C_{\rm {bd}}/2$. \cite{shib11}
In this study, $C_0=0$ and $k_{\rm {bend}}=k_{\rm{tilt}}=10$ are used except for the membrane particles belonging to the protein rods.

An anisotropic protein and membrane underneath it are together modeled as a rod
that is a linear chain of $N_{\rm {sg}}$ membrane particles~\cite{nogu14}.
We use $N_{\rm {sg}}=5$ and $10$ with
the density $\phi= N_{\rm {sg}}N_{\rm {rod}}/N= 0.167$.
The protein rods have spontaneous curvatures $C_{\rm {rod}}$ along the rod axis
and have no spontaneous (side) curvatures perpendicular to the rod axis.
The protein-bound membrane are more rigid than the bare membrane:
the values of $k_{\rm {bend}}$ and $k_{\rm{tilt}}$ are $k_{\rm r}$ times higher than those of  the bare membrane.

The membrane has mechanical properties that are typical of lipid membranes:
the bare membrane has a bending rigidity $\kappa/k_{\rm B}T=16.1 \pm 0.02$,
area of the tensionless membrane per particle $a_0/\sigma^2=1.2778\pm 0.0002$,
area compression modulus $K_A\sigma^2/k_{\rm B}T=83.1 \pm 0.4$,
 edge line tension $\Gamma\sigma/k_{\rm B}T= 5.73 \pm 0.04$~\cite{nogu14},
and the Gaussian modulus $\bar{\kappa}/\kappa=-0.9\pm 0.1$~\cite{nogu19}.
The bending rigidity is calculated by Eq.~(\ref{eq:fmb}),
which is slightly greater than the value ($15 \pm 1$) estimated by thermal undulation~\cite{shib11}.
The membrane tube with a length of $L_{\rm cy}$ is connected by the periodic boundary,
and the tube volume can be freely varied.
Molecular dynamics with a Langevin thermostat is employed~\cite{shib11,nogu11}.
The dependence on the rod curvature $C_{\rm rod}$ was calculated at $L_{\rm cy}=48\sigma$ and $N=2400$
 in Ref.~\citenum{nogu16a} using the replica-exchange method \cite{huku96,okam04}.
The dependence on the tube radius was calculated at $k_{\rm r}=4$ and $N=4800$ in this study.

\begin{figure}
\includegraphics[]{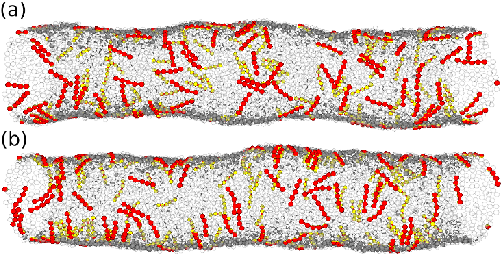}
\includegraphics[]{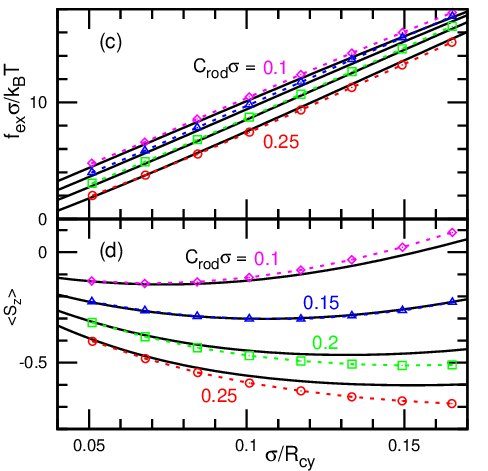}
\caption{
Membrane simulations of the short protein rods of $N_{\rm sg} = 5$ at $k_{\rm r}=4$.  
(a),(b) Snapshots for (a) $C_{\rm rod}\sigma = 0.1$ and (b) $C_{\rm rod}\sigma = 0.25$ 
at $L_{\rm cy}/\sigma=96$ ($R_{\rm cy}/\sigma=9.92$ and $9.91$, respectively).
A protein rod is displayed as
a chain of spheres whose halves are colored
in red and in yellow.
The orientational vector ${\bf u}_i$ lies along the direction from the 
yellow to red hemispheres.
Transparent gray particles represent
membrane particles.
(c),(d) Dependence of (c) the axial force $f_{\rm ex}$ and (d) orientational order $\langle S_z\rangle$ along the membrane tube 
on the tube radius $R_{\rm cy}$ for $C_{\rm rod}\sigma = 0.1$, $0.15$, $0.2$, and $0.25$.
The symbols with dashed lines represent the simulation data.
The black solid lines represent the theoretical results.
}
\label{fig:sim5}
\end{figure}

\begin{figure}
\includegraphics[]{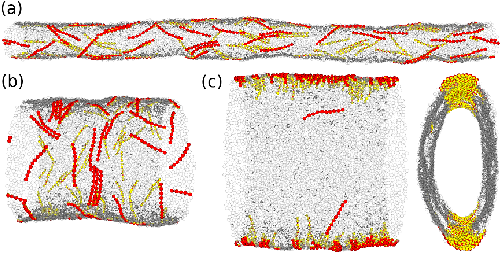}
\includegraphics[]{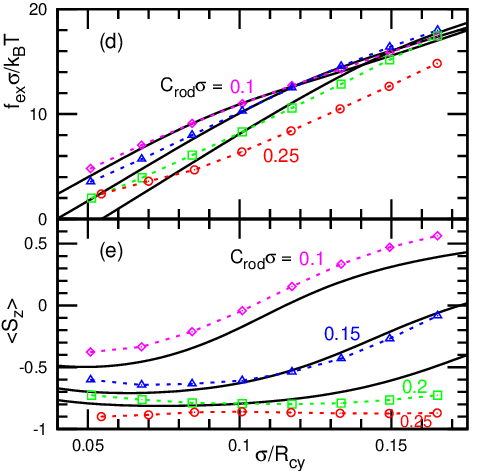}
\caption{
Membrane simulations of the long protein rods of $N_{\rm sg} = 10$ at $k_{\rm r}=4$.  
(a)--(c) Snapshots for (a),(b) $C_{\rm rod}\sigma = 0.1$ and (c) $C_{\rm rod}\sigma = 0.25$.
(a) $L_{\rm cy}/\sigma=160$ ($R_{\rm cy}/\sigma=6.05$).
(b),(c) $L_{\rm cy}/\sigma=48$ ($R_{\rm cy}/\sigma= 19.66$ and $18.37$, respectively).
The front and side views are displayed in (c).
(d),(e) Dependence of (d) the axial force $f_{\rm ex}$ and (e) orientational order $\langle S_z\rangle$ along the membrane tube 
on the tube radius $R_{\rm cy}$ for $C_{\rm rod}\sigma = 0.1$, $0.15$, $0.2$, and $0.25$.
The symbols with dashed lines represent the simulation data.
The black solid lines represent the theoretical results for $C_{\rm rod}\sigma = 0.1$, $0.15$, and $0.2$.
}
\label{fig:sim10}
\end{figure}

\begin{figure}
\includegraphics[]{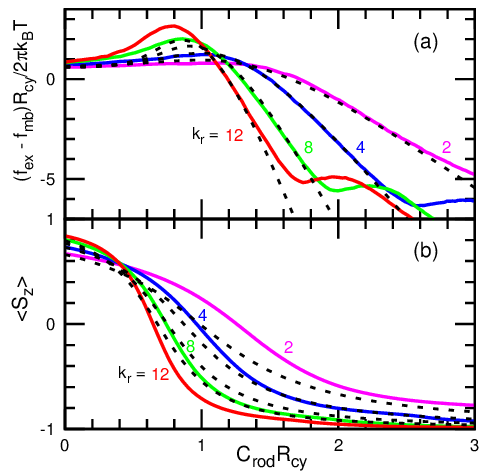}
\caption{
(a),(b) Dependence of (a) the axial force $f_{\rm ex}$ and (b) orientational order $\langle S_z\rangle$ along the membrane tube 
on the rod curvature $C_{\rm rod}$ for the bending rigidity ratio $k_{\rm r} = 2$, $4$, $8$, and $12$ at  $N_{\rm sg} = 10$ and  $R_{\rm cy}/\sigma=9.89$.
The solid lines represent the simulation data.
The black dashed lines represent the theoretical results.
The force generated by the proteins is normalized as
$(f_{\rm ex}-f_{\rm mb})R_{\rm cy}/2\pi k_{\rm B}T$, where $f_{\rm mb}$ is the force of the bare membrane.
The simulation data (solid lines) are reproduced from  Ref.~\citenum{nogu16a}.
}
\label{fig:cylvir}
\end{figure}

\subsection{Comparison of simulation and theoretical results}\label{sec:sres}

Figure~\ref{fig:sim5} and Figs.~\ref{fig:sim10},\ref{fig:cylvir} show
the simulation and theoretical results for the short and long protein rods ($N_{\rm sg} = 5$ and $10$), respectively.
Since the simulated proteins do not have an elliptic shape and are flexible,
the protein parameters are adjusted as follows.
For the short rods, we used
the orientational degree $S_z$ at $C_{\rm rod}\sigma=0.15$ (the second line from the top in Fig.~\ref{fig:sim5}(d)) for a fit
and obtained $\kappa_{\rm p}=30k_{\rm B}T$ and
 $C_{\rm rod}/C_{\rm p}=2$ for $a_{\rm p}=N_{\rm sg}a_{0}$ and $d_{\rm el}=3$.
This parameter set reproduces the simulation data of $S_z$ and $f_{\rm ex}$ at different values of $C_{\rm rod}$ very well.
Thus, this theory can quantitatively describe the behavior of the short proteins.

However, less agreement is obtained for the long rods of $N_{\rm sg} = 10$ (see Figs.~\ref{fig:sim10} and \ref{fig:cylvir}).
It is due to the protein assembly induced by the membrane-mediated attractive interactions 
between the proteins (see the snapshots in Figs.~\ref{fig:sim10}(a)--(c)).
At a high rod curvature ($C_{\rm rod}\sigma=0.25$), the proteins assemble in the azimuthal direction,
and the membrane deforms into an elliptic tube, as shown in Fig.~\ref{fig:sim10}(c).
For longer (narrower) and shorter (wider) tubes, cylindrical and triangular shapes are formed (see Movie 1 in ESI).
Thus, large negative values of $S_z$ (Fig.~\ref{fig:sim10}(e)) and nonmonotonic dependence of $f_{\rm ex}$ for $C_{\rm rod}R_{\rm cy} \gtrsim 2.5$
 ($C_{\rm rod}\sigma \gtrsim 0.25$) at $k_{\rm r}=4$ (Fig.~\ref{fig:cylvir}(a)) are obtained. 
In the  elliptic and triangular membranes, the proteins align in the azimuthal direction,
so that  their stabilities can be analyzed by assuming a fixed protein orientation as reported in Ref.~\citenum{nogu15b}.
More detail of this assembly is described in Refs.~\citenum{nogu14,nogu15b,nogu16a}.

For a lower rod curvature ($C_{\rm rod}\sigma \leq 0.2$) of the long rods at $k_{\rm r}=4$, the azimuthal assembly does not occur,
but clusters of a few proteins appear as shown in Figs.~\ref{fig:sim10}(a) and (b).
We fitted the linear-decrease region of the force-dependence curve in Fig.~\ref{fig:cylvir}(a) at $C_{\rm rod}R_{\rm cy}>1$
and obtained $\kappa_{\rm p}/k_{\rm B}T=60$, $90$, $120$, and $150$ with
 $C_{\rm rod}/C_{\rm p}=2.5$, $2.05$, $1.7$, and $1.5$ for $k_{\rm r}= 2$, $4$, $8$, and $12$, respectively, 
at $a_{\rm p}=N_{\rm sg}a_{0}$ and $d_{\rm el}=7$.
 The orientational orders $S_z$ calculated by these parameter sets
show quantitative deviation from the simulation data,
although they capture qualitative behavior (see Figs.~\ref{fig:sim10}(e) and \ref{fig:cylvir}(b)).
Moreover, the other regions of the force-dependence curves have quantitative differences:
the heights of peaks at  $C_{\rm rod}R_{\rm cy}<1$ in Fig.~\ref{fig:cylvir}(a) 
deviate from the simulation values,
and the slopes at  $\sigma/R_{\rm cy}<0.1$  in Fig.~\ref{fig:sim10}(d) are different.
Although the present theory assumes the uniform lateral distribution of the proteins,
the protein clusters can bend the membrane more strongly as demonstrated by the formation of the elliptic tube.
Therefore, we consider that the clusters effectively work as large or rigid proteins.
The greater values of $\kappa_{\rm p}$ and $C_{\rm p}$ obtained by the fits
support this mechanism.
Thus, for a quantitative prediction of a long protein (i.e., a large elliptic ratio $d_{\rm el}$),
it is significant to include the effects of the protein clusters.

\section{Summary and discussions}\label{sec:sum}

We have studied the equilibrium states of the anisotropic curvature-inducing proteins theoretically 
and compared them with the simulation results.
The protein is assumed to have an elliptic shape with a bending rigidity and spontaneous curvature
mainly along the major axis of the protein.
On narrow membrane tubes, the proteins exhibit a first-order nematic transition with increasing protein density
as reported in our previous paper~\cite{tozz21}.
Here, we found that this transition becomes the second order on the tubes with intermediate radii.
In our previous study, the proteins on a membrane with a fixed shape have been considered.
In this study, we extended the theory to proteins on membrane tubes which radius is not fixed
and in the binding/unbinding equilibrium.
We found that the protein binding affects the membrane axial force differently for wide and narrow tubes.
For wide tubes, the force is reduced by the binding.
In contrast, it is only slightly modified for narrow tubes, on which the proteins are tilted from the azimuthal direction.
With increasing binding chemical potential,
a first-order transition between two tube radii with different protein densities
occurs only once at the wide tubes, whereas the proteins with an isotropic bending energy
exhibit the transition twice.
For the short proteins,
this theory reproduces the protein orientation and axial force obtained by the meshless simulations very well.
In contrast, the long proteins have large membrane-mediated attractive interactions
so that resultant protein clusters modify the mean orientation and axial force.
However, the theory still holds qualitative dependency. 

Moreover, we found that the tube curvatures for the maximum protein binding (sensing) and orientational order
are different from the protein spontaneous curvature $C_{\rm p}$.
The sensing curvature is higher than $C_{\rm p}$ at low protein density
and coincides to the curvature of the free energy minimum.
This is contrast to isotropic proteins~\cite{nogu21b} which sensing curvature is constant.
The order curvature is lower than  $C_{\rm p}$ and decreases with increasing protein density and bending rigidity.
These dependencies are caused by the variation in the protein orientation.
Previously, the proteins are often assumed to orient to the azimuthal direction.
Even at the tube curvature close to $C_{\rm p}$, the orientational fluctuations modify the average protein behavior.
Thus, it is important to take the orientational degree of freedom into account.

Since the theory for the isotropic bending energy has been well established,
it has been employed even in the analysis for the experiments of the BAR proteins~\cite{prev15,wu18,tsai21}.
In this study, however, we have clarified that the anisotropy of the bending energy largely changes the membrane--protein interactions
such as the sensing curvature.
Therefore, the effects of the protein orientation should be included for
more quantitative analysis.

We also showed that the protein side curvature (spontaneous curvature along the protein minor axis)
 modifies the protein binding.
The proteins are oriented less or more strongly in the azimuthal direction 
for a positive or negative side curvature, respectively.
It has been reported that
 the side curvature of the opposite sign to $C_{\rm p}$ can induce
 the formation of an egg-carton shape~\cite{domm99,domm02} and  network structures~\cite{nogu16}.
However, in many previous studies, the side curvature has not been considered.
When the proteins or objects strongly bind the membrane 
as in coarse-grained molecular simulations~\cite{simu13,olin16},
the proteins effectively have a large negative side curvature.
Such a large side curvature can change the orientation direction perpendicularly
leading to the tip-to-tip protein assembly as discussed in Ref.~\citenum{nogu17}.

In the present theory, we consider only an excluded-volume interaction between proteins.
Bound proteins can attract each other via direct and/or
 membrane-mediated interactions.
In particular, BAR proteins typically form helical alignments in dense-packed conditions.
Such a chiral interaction can largely modify the protein assembly and membrane shape~\cite{nogu19a,behe21}.
To reproduce them, additional interactions are required to account for.
However, for a sufficiently low protein density,
such additional interactions are negligibly weak.
Hence, 
the present theory  can be used to estimate the bending rigidity and curvature of bound proteins
in experiments and atomistic simulations.
These mechanical parameters are keys to quantitatively understand the curvature sensing and generation of proteins.

\begin{acknowledgments}
This work was supported by JSPS KAKENHI Grant Number JP21K03481,
the European Research Council (CoG-681434), the European Commission (Project No. H2020-FETPROACT-01-2016-731957), the Spanish Ministry for Science and Innovation/FEDER (PID2019-110949GBI00, BES-2016-078220 to C. T.), and the Generalitat de Catalunya (ICREA Academia award).
This research was also supported in part by the National Science Foundation under Grant No. NSF PHY-1748958, through KITP program: The Physics of Elastic Films: from Biological Membranes to Extreme Mechanics (FILMS21).
The simulations were partially carried out using the facilities of the Supercomputer Center, Institute for Solid State Physics, University of Tokyo. 
\end{acknowledgments}

\begin{appendix}

\section{Calculation method for the present theory}\label{sec:psi}

The quantities $\Psi$ and $\bar{\Psi}$ are
single-valued functions of $S$,
so that they can be calculated by root-finding algorithms such as the bisection method.
Alternatively, for $\phi \leq 1/(b_0 + b_2/2)$, where $g>0$ for any value of $S$,
$S$ can be calculated by solving the quadratic equation of 
$S= (q_0 + q_1 S)/(p_0+ p_1 S)$ for given $\Psi$ and $\bar{\Psi}$  as
\begin{equation}
S = \frac{q_1 - p_0 + \sqrt{(q_1-p_0)^2 + 4q_0p_1 } }{2 p_1}, 
\end{equation}
where
\begin{eqnarray}
p_0 &=& \int_{-\pi/2}^{\pi/2}\ w_0 \ d\theta_{\rm ps},
\hspace{1.05cm} p_1 = \int_{-\pi/2}^{\pi/2}\ w_1 \ d\theta_{\rm ps},  \nonumber \\
q_0 &=& \int_{-\pi/2}^{\pi/2} 2s_{\rm p}(\theta_{\rm ps}) w_0 \ d\theta_{\rm ps}, 
\hspace{0.2cm} q_1 = \int_{-\pi/2}^{\pi/2} 2s_{\rm p}(\theta_{\rm ps}) w_1 \ d\theta_{\rm ps},  \nonumber \\
w_0 &=& (1-b_0\phi)\exp[\Psi s_{\rm p}(\theta_{\rm ps}) + \bar{\Psi}\sin(\theta_{\rm ps})\cos(\theta_{\rm ps}) - \beta U_{\rm p} ],  \nonumber \\ \nonumber
w_1 &=& b_2\phi s_{\rm p}(\theta_{\rm ps})\exp[\Psi s_{\rm p}(\theta_{\rm ps}) + \bar{\Psi}\sin(\theta_{\rm ps})\cos(\theta_{\rm ps}) - \beta U_{\rm p} ]. 
\end{eqnarray}
At $\phi \simeq 1/b_0$, $S$ can be a multivalued function of $\Psi$.
For $1/(b_0 + b_2/2) < \phi < \phi_{\rm max}$,
large values of $S$ can be calculated by the iteration of Eq.~(\ref{eq:s}) by updating $S$. 
However, the smaller values of $S$ should be calculated by the former method ($\Psi(S)$).
In this study, $\partial f_{\rm p}/\partial (1/R_{\rm cy})$ and
$\partial f_{\rm p}/\partial \phi$ are calculated by the central difference method.

\end{appendix}


\begin{thebibliography}{70}%
\makeatletter
\providecommand \@ifxundefined [1]{%
 \@ifx{#1\undefined}
}%
\providecommand \@ifnum [1]{%
 \ifnum #1\expandafter \@firstoftwo
 \else \expandafter \@secondoftwo
 \fi
}%
\providecommand \@ifx [1]{%
 \ifx #1\expandafter \@firstoftwo
 \else \expandafter \@secondoftwo
 \fi
}%
\providecommand \natexlab [1]{#1}%
\providecommand \enquote  [1]{``#1''}%
\providecommand \bibnamefont  [1]{#1}%
\providecommand \bibfnamefont [1]{#1}%
\providecommand \citenamefont [1]{#1}%
\providecommand \href@noop [0]{\@secondoftwo}%
\providecommand \href [0]{\begingroup \@sanitize@url \@href}%
\providecommand \@href[1]{\@@startlink{#1}\@@href}%
\providecommand \@@href[1]{\endgroup#1\@@endlink}%
\providecommand \@sanitize@url [0]{\catcode `\\12\catcode `\$12\catcode
  `\&12\catcode `\#12\catcode `\^12\catcode `\_12\catcode `\%12\relax}%
\providecommand \@@startlink[1]{}%
\providecommand \@@endlink[0]{}%
\providecommand \url  [0]{\begingroup\@sanitize@url \@url }%
\providecommand \@url [1]{\endgroup\@href {#1}{\urlprefix }}%
\providecommand \urlprefix  [0]{URL }%
\providecommand \Eprint [0]{\href }%
\providecommand \doibase [0]{http://dx.doi.org/}%
\providecommand \selectlanguage [0]{\@gobble}%
\providecommand \bibinfo  [0]{\@secondoftwo}%
\providecommand \bibfield  [0]{\@secondoftwo}%
\providecommand \translation [1]{[#1]}%
\providecommand \BibitemOpen [0]{}%
\providecommand \bibitemStop [0]{}%
\providecommand \bibitemNoStop [0]{.\EOS\space}%
\providecommand \EOS [0]{\spacefactor3000\relax}%
\providecommand \BibitemShut  [1]{\csname bibitem#1\endcsname}%
\let\auto@bib@innerbib\@empty
\bibitem [{\citenamefont {McMahon}\ and\ \citenamefont
  {Gallop}(2005)}]{mcma05}%
  \BibitemOpen
  \bibfield  {author} {\bibinfo {author} {\bibfnamefont {H.~T.}\ \bibnamefont
  {McMahon}}\ and\ \bibinfo {author} {\bibfnamefont {J.~L.}\ \bibnamefont
  {Gallop}},\ }\href@noop {} {\bibfield  {journal} {\bibinfo  {journal}
  {Nature}\ }\textbf {\bibinfo {volume} {438}},\ \bibinfo {pages} {590}
  (\bibinfo {year} {2005})}\BibitemShut {NoStop}%
\bibitem [{\citenamefont {Suetsugu}\ \emph {et~al.}(2014)\citenamefont
  {Suetsugu}, \citenamefont {Kurisu},\ and\ \citenamefont {Takenawa}}]{suet14}%
  \BibitemOpen
  \bibfield  {author} {\bibinfo {author} {\bibfnamefont {S.}~\bibnamefont
  {Suetsugu}}, \bibinfo {author} {\bibfnamefont {S.}~\bibnamefont {Kurisu}}, \
  and\ \bibinfo {author} {\bibfnamefont {T.}~\bibnamefont {Takenawa}},\
  }\href@noop {} {\bibfield  {journal} {\bibinfo  {journal} {Physiol. Rev.}\
  }\textbf {\bibinfo {volume} {94}},\ \bibinfo {pages} {1219} (\bibinfo {year}
  {2014})}\BibitemShut {NoStop}%
\bibitem [{\citenamefont {Johannes}\ \emph {et~al.}(2015)\citenamefont
  {Johannes}, \citenamefont {Parton}, \citenamefont {Bassereau},\ and\
  \citenamefont {Mayor}}]{joha15}%
  \BibitemOpen
  \bibfield  {author} {\bibinfo {author} {\bibfnamefont {L.}~\bibnamefont
  {Johannes}}, \bibinfo {author} {\bibfnamefont {R.~G.}\ \bibnamefont
  {Parton}}, \bibinfo {author} {\bibfnamefont {P.}~\bibnamefont {Bassereau}}, \
  and\ \bibinfo {author} {\bibfnamefont {S.}~\bibnamefont {Mayor}},\
  }\href@noop {} {\bibfield  {journal} {\bibinfo  {journal} {Nat. Rev. Mol.
  Cell. Biol.}\ }\textbf {\bibinfo {volume} {16}},\ \bibinfo {pages} {311}
  (\bibinfo {year} {2015})}\BibitemShut {NoStop}%
\bibitem [{\citenamefont {Brandizzi}\ and\ \citenamefont
  {Barlowe}(2013)}]{bran13}%
  \BibitemOpen
  \bibfield  {author} {\bibinfo {author} {\bibfnamefont {F.}~\bibnamefont
  {Brandizzi}}\ and\ \bibinfo {author} {\bibfnamefont {C.}~\bibnamefont
  {Barlowe}},\ }\href@noop {} {\bibfield  {journal} {\bibinfo  {journal} {Nat.
  Rev. Mol. Cell Biol.}\ }\textbf {\bibinfo {volume} {14}},\ \bibinfo {pages}
  {382} (\bibinfo {year} {2013})}\BibitemShut {NoStop}%
\bibitem [{\citenamefont {Hurley}\ \emph {et~al.}(2010)\citenamefont {Hurley},
  \citenamefont {Boura}, \citenamefont {Carlson},\ and\ \citenamefont
  {R{\'o}{\.{z}}ycki}}]{hurl10}%
  \BibitemOpen
  \bibfield  {author} {\bibinfo {author} {\bibfnamefont {J.~H.}\ \bibnamefont
  {Hurley}}, \bibinfo {author} {\bibfnamefont {E.}~\bibnamefont {Boura}},
  \bibinfo {author} {\bibfnamefont {L.-A.}\ \bibnamefont {Carlson}}, \ and\
  \bibinfo {author} {\bibfnamefont {B.}~\bibnamefont {R{\'o}{\.{z}}ycki}},\
  }\href@noop {} {\bibfield  {journal} {\bibinfo  {journal} {Cell}\ }\textbf
  {\bibinfo {volume} {143}},\ \bibinfo {pages} {875} (\bibinfo {year}
  {2010})}\BibitemShut {NoStop}%
\bibitem [{\citenamefont {McMahon}\ and\ \citenamefont
  {Boucrot}(2011)}]{mcma11}%
  \BibitemOpen
  \bibfield  {author} {\bibinfo {author} {\bibfnamefont {H.~T.}\ \bibnamefont
  {McMahon}}\ and\ \bibinfo {author} {\bibfnamefont {E.}~\bibnamefont
  {Boucrot}},\ }\href@noop {} {\bibfield  {journal} {\bibinfo  {journal} {Nat.
  Rev. Mol. Cell. Biol.}\ }\textbf {\bibinfo {volume} {12}},\ \bibinfo {pages}
  {517} (\bibinfo {year} {2011})}\BibitemShut {NoStop}%
\bibitem [{\citenamefont {Baumgart}\ \emph {et~al.}(2011)\citenamefont
  {Baumgart}, \citenamefont {Capraro}, \citenamefont {Zhu},\ and\ \citenamefont
  {Das}}]{baum11}%
  \BibitemOpen
  \bibfield  {author} {\bibinfo {author} {\bibfnamefont {T.}~\bibnamefont
  {Baumgart}}, \bibinfo {author} {\bibfnamefont {B.~R.}\ \bibnamefont
  {Capraro}}, \bibinfo {author} {\bibfnamefont {C.}~\bibnamefont {Zhu}}, \ and\
  \bibinfo {author} {\bibfnamefont {S.~L.}\ \bibnamefont {Das}},\ }\href@noop
  {} {\bibfield  {journal} {\bibinfo  {journal} {Annu. Rev. Phys. Chem.}\
  }\textbf {\bibinfo {volume} {62}},\ \bibinfo {pages} {483} (\bibinfo {year}
  {2011})}\BibitemShut {NoStop}%
\bibitem [{\citenamefont {Has}\ and\ \citenamefont {Das}(2021)}]{has21}%
  \BibitemOpen
  \bibfield  {author} {\bibinfo {author} {\bibfnamefont {C.}~\bibnamefont
  {Has}}\ and\ \bibinfo {author} {\bibfnamefont {S.~L.}\ \bibnamefont {Das}},\
  }\href {\doibase 10.1016/j.bbagen.2021.129971} {\bibfield  {journal}
  {\bibinfo  {journal} {Biochim.\ Biophys.\ Acta}\ }\textbf {\bibinfo {volume}
  {1865}},\ \bibinfo {pages} {129971} (\bibinfo {year} {2021})}\BibitemShut
  {NoStop}%
\bibitem [{\citenamefont {Itoh}\ and\ \citenamefont {{De
  Camilli}}(2006)}]{itoh06}%
  \BibitemOpen
  \bibfield  {author} {\bibinfo {author} {\bibfnamefont {T.}~\bibnamefont
  {Itoh}}\ and\ \bibinfo {author} {\bibfnamefont {P.}~\bibnamefont {{De
  Camilli}}},\ }\href@noop {} {\bibfield  {journal} {\bibinfo  {journal}
  {Biochim.\ Biophys.\ Acta}\ }\textbf {\bibinfo {volume} {1761}},\ \bibinfo
  {pages} {897} (\bibinfo {year} {2006})}\BibitemShut {NoStop}%
\bibitem [{\citenamefont {Masuda}\ and\ \citenamefont
  {Mochizuki}(2010)}]{masu10}%
  \BibitemOpen
  \bibfield  {author} {\bibinfo {author} {\bibfnamefont {M.}~\bibnamefont
  {Masuda}}\ and\ \bibinfo {author} {\bibfnamefont {N.}~\bibnamefont
  {Mochizuki}},\ }\href@noop {} {\bibfield  {journal} {\bibinfo  {journal}
  {Semin. Cell Dev. Biol.}\ }\textbf {\bibinfo {volume} {21}},\ \bibinfo
  {pages} {391} (\bibinfo {year} {2010})}\BibitemShut {NoStop}%
\bibitem [{\citenamefont {Mim}\ and\ \citenamefont {Unger}(2012)}]{mim12a}%
  \BibitemOpen
  \bibfield  {author} {\bibinfo {author} {\bibfnamefont {C.}~\bibnamefont
  {Mim}}\ and\ \bibinfo {author} {\bibfnamefont {V.~M.}\ \bibnamefont
  {Unger}},\ }\href@noop {} {\bibfield  {journal} {\bibinfo  {journal} {Trends
  Biochem. Sci.}\ }\textbf {\bibinfo {volume} {37}},\ \bibinfo {pages} {526}
  (\bibinfo {year} {2012})}\BibitemShut {NoStop}%
\bibitem [{\citenamefont {Frost}\ \emph {et~al.}(2008)\citenamefont {Frost},
  \citenamefont {Perera}, \citenamefont {Roux}, \citenamefont {Spasov},
  \citenamefont {Destaing}, \citenamefont {Egelman}, \citenamefont {{De
  Camilli}},\ and\ \citenamefont {Unger}}]{fros08}%
  \BibitemOpen
  \bibfield  {author} {\bibinfo {author} {\bibfnamefont {A.}~\bibnamefont
  {Frost}}, \bibinfo {author} {\bibfnamefont {R.}~\bibnamefont {Perera}},
  \bibinfo {author} {\bibfnamefont {A.}~\bibnamefont {Roux}}, \bibinfo {author}
  {\bibfnamefont {K.}~\bibnamefont {Spasov}}, \bibinfo {author} {\bibfnamefont
  {O.}~\bibnamefont {Destaing}}, \bibinfo {author} {\bibfnamefont {E.~H.}\
  \bibnamefont {Egelman}}, \bibinfo {author} {\bibfnamefont {P.}~\bibnamefont
  {{De Camilli}}}, \ and\ \bibinfo {author} {\bibfnamefont {V.~M.}\
  \bibnamefont {Unger}},\ }\href@noop {} {\bibfield  {journal} {\bibinfo
  {journal} {Cell}\ }\textbf {\bibinfo {volume} {132}},\ \bibinfo {pages} {807}
  (\bibinfo {year} {2008})}\BibitemShut {NoStop}%
\bibitem [{\citenamefont {Sorre}\ \emph {et~al.}(2012)\citenamefont {Sorre},
  \citenamefont {Callan-Jones}, \citenamefont {Manzi}, \citenamefont {Goud},
  \citenamefont {Prost}, \citenamefont {Bassereau},\ and\ \citenamefont
  {Roux}}]{sorr12}%
  \BibitemOpen
  \bibfield  {author} {\bibinfo {author} {\bibfnamefont {B.}~\bibnamefont
  {Sorre}}, \bibinfo {author} {\bibfnamefont {A.}~\bibnamefont {Callan-Jones}},
  \bibinfo {author} {\bibfnamefont {J.}~\bibnamefont {Manzi}}, \bibinfo
  {author} {\bibfnamefont {B.}~\bibnamefont {Goud}}, \bibinfo {author}
  {\bibfnamefont {J.}~\bibnamefont {Prost}}, \bibinfo {author} {\bibfnamefont
  {P.}~\bibnamefont {Bassereau}}, \ and\ \bibinfo {author} {\bibfnamefont
  {A.}~\bibnamefont {Roux}},\ }\href@noop {} {\bibfield  {journal} {\bibinfo
  {journal} {Proc.\ Natl.\ Acad.\ Sci.\ USA}\ }\textbf {\bibinfo {volume}
  {109}},\ \bibinfo {pages} {173} (\bibinfo {year} {2012})}\BibitemShut
  {NoStop}%
\bibitem [{\citenamefont {Zhu}\ \emph {et~al.}(2012)\citenamefont {Zhu},
  \citenamefont {Das},\ and\ \citenamefont {Baumgart}}]{zhu12}%
  \BibitemOpen
  \bibfield  {author} {\bibinfo {author} {\bibfnamefont {C.}~\bibnamefont
  {Zhu}}, \bibinfo {author} {\bibfnamefont {S.~L.}\ \bibnamefont {Das}}, \ and\
  \bibinfo {author} {\bibfnamefont {T.}~\bibnamefont {Baumgart}},\ }\href@noop
  {} {\bibfield  {journal} {\bibinfo  {journal} {Biophys. J.}\ }\textbf
  {\bibinfo {volume} {102}},\ \bibinfo {pages} {1837} (\bibinfo {year}
  {2012})}\BibitemShut {NoStop}%
\bibitem [{\citenamefont {Tanaka-Takiguchi}\ \emph {et~al.}(2013)\citenamefont
  {Tanaka-Takiguchi}, \citenamefont {Itoh}, \citenamefont {Tsujita},
  \citenamefont {Yamada}, \citenamefont {Yanagisawa}, \citenamefont {Fujiwara},
  \citenamefont {Yamamoto}, \citenamefont {Ichikawa},\ and\ \citenamefont
  {Takiguchi}}]{tana13}%
  \BibitemOpen
  \bibfield  {author} {\bibinfo {author} {\bibfnamefont {Y.}~\bibnamefont
  {Tanaka-Takiguchi}}, \bibinfo {author} {\bibfnamefont {T.}~\bibnamefont
  {Itoh}}, \bibinfo {author} {\bibfnamefont {K.}~\bibnamefont {Tsujita}},
  \bibinfo {author} {\bibfnamefont {S.}~\bibnamefont {Yamada}}, \bibinfo
  {author} {\bibfnamefont {M.}~\bibnamefont {Yanagisawa}}, \bibinfo {author}
  {\bibfnamefont {K.}~\bibnamefont {Fujiwara}}, \bibinfo {author}
  {\bibfnamefont {A.}~\bibnamefont {Yamamoto}}, \bibinfo {author}
  {\bibfnamefont {M.}~\bibnamefont {Ichikawa}}, \ and\ \bibinfo {author}
  {\bibfnamefont {K.}~\bibnamefont {Takiguchi}},\ }\href@noop {} {\bibfield
  {journal} {\bibinfo  {journal} {Langmuir}\ }\textbf {\bibinfo {volume}
  {29}},\ \bibinfo {pages} {328} (\bibinfo {year} {2013})}\BibitemShut
  {NoStop}%
\bibitem [{\citenamefont {Adam}\ \emph {et~al.}(2015)\citenamefont {Adam},
  \citenamefont {Basnet},\ and\ \citenamefont {Mizuno}}]{adam15}%
  \BibitemOpen
  \bibfield  {author} {\bibinfo {author} {\bibfnamefont {J.}~\bibnamefont
  {Adam}}, \bibinfo {author} {\bibfnamefont {N.}~\bibnamefont {Basnet}}, \ and\
  \bibinfo {author} {\bibfnamefont {N.}~\bibnamefont {Mizuno}},\ }\href@noop {}
  {\bibfield  {journal} {\bibinfo  {journal} {Sci. Rep.}\ }\textbf {\bibinfo
  {volume} {5}},\ \bibinfo {pages} {15452} (\bibinfo {year}
  {2015})}\BibitemShut {NoStop}%
\bibitem [{\citenamefont {Pr{\'e}vost}\ \emph {et~al.}(2015)\citenamefont
  {Pr{\'e}vost}, \citenamefont {Zhao}, \citenamefont {Manzi}, \citenamefont
  {Lemichez}, \citenamefont {Lappalainen}, \citenamefont {Callan-Jones},\ and\
  \citenamefont {Bassereau}}]{prev15}%
  \BibitemOpen
  \bibfield  {author} {\bibinfo {author} {\bibfnamefont {C.}~\bibnamefont
  {Pr{\'e}vost}}, \bibinfo {author} {\bibfnamefont {H.}~\bibnamefont {Zhao}},
  \bibinfo {author} {\bibfnamefont {J.}~\bibnamefont {Manzi}}, \bibinfo
  {author} {\bibfnamefont {E.}~\bibnamefont {Lemichez}}, \bibinfo {author}
  {\bibfnamefont {P.}~\bibnamefont {Lappalainen}}, \bibinfo {author}
  {\bibfnamefont {A.}~\bibnamefont {Callan-Jones}}, \ and\ \bibinfo {author}
  {\bibfnamefont {P.}~\bibnamefont {Bassereau}},\ }\href {\doibase
  10.1038/ncomms9529} {\bibfield  {journal} {\bibinfo  {journal} {Nat.
  Commun.}\ }\textbf {\bibinfo {volume} {6}},\ \bibinfo {pages} {8529}
  (\bibinfo {year} {2015})}\BibitemShut {NoStop}%
\bibitem [{\citenamefont {Tsai}\ \emph {et~al.}(2021)\citenamefont {Tsai},
  \citenamefont {Simunovic}, \citenamefont {Sorre}, \citenamefont {Bertin},
  \citenamefont {Manzi}, \citenamefont {Callan-Jones},\ and\ \citenamefont
  {Bassereau}}]{tsai21}%
  \BibitemOpen
  \bibfield  {author} {\bibinfo {author} {\bibfnamefont {F.-C.}\ \bibnamefont
  {Tsai}}, \bibinfo {author} {\bibfnamefont {M.}~\bibnamefont {Simunovic}},
  \bibinfo {author} {\bibfnamefont {B.}~\bibnamefont {Sorre}}, \bibinfo
  {author} {\bibfnamefont {A.}~\bibnamefont {Bertin}}, \bibinfo {author}
  {\bibfnamefont {J.}~\bibnamefont {Manzi}}, \bibinfo {author} {\bibfnamefont
  {A.}~\bibnamefont {Callan-Jones}}, \ and\ \bibinfo {author} {\bibfnamefont
  {P.}~\bibnamefont {Bassereau}},\ }\href {\doibase 10.1039/d0sm01573c}
  {\bibfield  {journal} {\bibinfo  {journal} {Soft Matter}\ }\textbf {\bibinfo
  {volume} {17}},\ \bibinfo {pages} {4254} (\bibinfo {year}
  {2021})}\BibitemShut {NoStop}%
\bibitem [{\citenamefont {Roux}\ \emph {et~al.}(2010)\citenamefont {Roux},
  \citenamefont {Koster}, \citenamefont {Lenz}, \citenamefont {Sorre},
  \citenamefont {Manneville}, \citenamefont {Nassoy},\ and\ \citenamefont
  {Bassereau}}]{roux10}%
  \BibitemOpen
  \bibfield  {author} {\bibinfo {author} {\bibfnamefont {A.}~\bibnamefont
  {Roux}}, \bibinfo {author} {\bibfnamefont {G.}~\bibnamefont {Koster}},
  \bibinfo {author} {\bibfnamefont {M.}~\bibnamefont {Lenz}}, \bibinfo {author}
  {\bibfnamefont {B.}~\bibnamefont {Sorre}}, \bibinfo {author} {\bibfnamefont
  {J.-B.}\ \bibnamefont {Manneville}}, \bibinfo {author} {\bibfnamefont
  {P.}~\bibnamefont {Nassoy}}, \ and\ \bibinfo {author} {\bibfnamefont
  {P.}~\bibnamefont {Bassereau}},\ }\href@noop {} {\bibfield  {journal}
  {\bibinfo  {journal} {Proc.\ Natl.\ Acad.\ Sci.\ USA}\ }\textbf {\bibinfo
  {volume} {107}},\ \bibinfo {pages} {4141} (\bibinfo {year}
  {2010})}\BibitemShut {NoStop}%
\bibitem [{\citenamefont {Rosholm}\ \emph {et~al.}(2017)\citenamefont
  {Rosholm}, \citenamefont {Leijnse}, \citenamefont {Mantsiou}, \citenamefont
  {Tkach}, \citenamefont {Pedersen}, \citenamefont {Wirth}, \citenamefont
  {Oddershede}, \citenamefont {Jensen}, \citenamefont {Martinez}, \citenamefont
  {Hatzakis}, \citenamefont {Bendix}, \citenamefont {Callan-Jones},\ and\
  \citenamefont {Stamou}}]{rosh17}%
  \BibitemOpen
  \bibfield  {author} {\bibinfo {author} {\bibfnamefont {K.~R.}\ \bibnamefont
  {Rosholm}}, \bibinfo {author} {\bibfnamefont {N.}~\bibnamefont {Leijnse}},
  \bibinfo {author} {\bibfnamefont {A.}~\bibnamefont {Mantsiou}}, \bibinfo
  {author} {\bibfnamefont {V.}~\bibnamefont {Tkach}}, \bibinfo {author}
  {\bibfnamefont {S.~L.}\ \bibnamefont {Pedersen}}, \bibinfo {author}
  {\bibfnamefont {V.~F.}\ \bibnamefont {Wirth}}, \bibinfo {author}
  {\bibfnamefont {L.~B.}\ \bibnamefont {Oddershede}}, \bibinfo {author}
  {\bibfnamefont {K.~J.}\ \bibnamefont {Jensen}}, \bibinfo {author}
  {\bibfnamefont {K.~L.}\ \bibnamefont {Martinez}}, \bibinfo {author}
  {\bibfnamefont {N.~S.}\ \bibnamefont {Hatzakis}}, \bibinfo {author}
  {\bibfnamefont {P.~M.}\ \bibnamefont {Bendix}}, \bibinfo {author}
  {\bibfnamefont {A.}~\bibnamefont {Callan-Jones}}, \ and\ \bibinfo {author}
  {\bibfnamefont {D.}~\bibnamefont {Stamou}},\ }\href@noop {} {\bibfield
  {journal} {\bibinfo  {journal} {Nat. Chem. Biol.}\ }\textbf {\bibinfo
  {volume} {13}},\ \bibinfo {pages} {724} (\bibinfo {year} {2017})}\BibitemShut
  {NoStop}%
\bibitem [{\citenamefont {Canham}(1970)}]{canh70}%
  \BibitemOpen
  \bibfield  {author} {\bibinfo {author} {\bibfnamefont {P.~B.}\ \bibnamefont
  {Canham}},\ }\href@noop {} {\bibfield  {journal} {\bibinfo  {journal} {J.
  Theor. Biol.}\ }\textbf {\bibinfo {volume} {26}},\ \bibinfo {pages} {61}
  (\bibinfo {year} {1970})}\BibitemShut {NoStop}%
\bibitem [{\citenamefont {Helfrich}(1973)}]{helf73}%
  \BibitemOpen
  \bibfield  {author} {\bibinfo {author} {\bibfnamefont {W.}~\bibnamefont
  {Helfrich}},\ }\href@noop {} {\bibfield  {journal} {\bibinfo  {journal} {Z.\
  Naturforsch}\ }\textbf {\bibinfo {volume} {28c}},\ \bibinfo {pages} {693}
  (\bibinfo {year} {1973})}\BibitemShut {NoStop}%
\bibitem [{\citenamefont {Lipowsky}(1992)}]{lipo92}%
  \BibitemOpen
  \bibfield  {author} {\bibinfo {author} {\bibfnamefont {R.}~\bibnamefont
  {Lipowsky}},\ }\href@noop {} {\bibfield  {journal} {\bibinfo  {journal} {J.
  Phys. II France}\ }\textbf {\bibinfo {volume} {2}},\ \bibinfo {pages} {1825}
  (\bibinfo {year} {1992})}\BibitemShut {NoStop}%
\bibitem [{\citenamefont {Sens}(2004)}]{sens03}%
  \BibitemOpen
  \bibfield  {author} {\bibinfo {author} {\bibfnamefont {P.}~\bibnamefont
  {Sens}},\ }\href@noop {} {\bibfield  {journal} {\bibinfo  {journal} {Phys.
  Rev. Lett.}\ }\textbf {\bibinfo {volume} {93}},\ \bibinfo {pages} {108103}
  (\bibinfo {year} {2004})}\BibitemShut {NoStop}%
\bibitem [{\citenamefont {Foret}(2014)}]{fore14}%
  \BibitemOpen
  \bibfield  {author} {\bibinfo {author} {\bibfnamefont {L.}~\bibnamefont
  {Foret}},\ }\href@noop {} {\bibfield  {journal} {\bibinfo  {journal} {Eur.
  Phys. J. E}\ }\textbf {\bibinfo {volume} {37}},\ \bibinfo {pages} {42}
  (\bibinfo {year} {2014})}\BibitemShut {NoStop}%
\bibitem [{\citenamefont {Frey}\ and\ \citenamefont {Schwarz}(2020)}]{frey20}%
  \BibitemOpen
  \bibfield  {author} {\bibinfo {author} {\bibfnamefont {F.}~\bibnamefont
  {Frey}}\ and\ \bibinfo {author} {\bibfnamefont {U.~S.}\ \bibnamefont
  {Schwarz}},\ }\href@noop {} {\bibfield  {journal} {\bibinfo  {journal} {Soft
  Matter}\ }\textbf {\bibinfo {volume} {16}},\ \bibinfo {pages} {10723}
  (\bibinfo {year} {2020})}\BibitemShut {NoStop}%
\bibitem [{\citenamefont {Tozzi}\ \emph {et~al.}(2019)\citenamefont {Tozzi},
  \citenamefont {Walani},\ and\ \citenamefont {Arroyo}}]{tozz19}%
  \BibitemOpen
  \bibfield  {author} {\bibinfo {author} {\bibfnamefont {C.}~\bibnamefont
  {Tozzi}}, \bibinfo {author} {\bibfnamefont {N.}~\bibnamefont {Walani}}, \
  and\ \bibinfo {author} {\bibfnamefont {M.}~\bibnamefont {Arroyo}},\ }\href
  {\doibase 10.1088/1367-2630/ab3ad6} {\bibfield  {journal} {\bibinfo
  {journal} {New J. Phys.}\ }\textbf {\bibinfo {volume} {21}},\ \bibinfo
  {pages} {093004} (\bibinfo {year} {2019})}\BibitemShut {NoStop}%
\bibitem [{\citenamefont {Noguchi}(2021{\natexlab{a}})}]{nogu21a}%
  \BibitemOpen
  \bibfield  {author} {\bibinfo {author} {\bibfnamefont {H.}~\bibnamefont
  {Noguchi}},\ }\href {\doibase 10.1103/PhysRevE.104.014410} {\bibfield
  {journal} {\bibinfo  {journal} {Phys. Rev. E}\ }\textbf {\bibinfo {volume}
  {104}},\ \bibinfo {pages} {014410} (\bibinfo {year}
  {2021}{\natexlab{a}})}\BibitemShut {NoStop}%
\bibitem [{\citenamefont {Goutaland}\ \emph {et~al.}(2021)\citenamefont
  {Goutaland}, \citenamefont {van Wijland}, \citenamefont {Fournier},\ and\
  \citenamefont {Noguchi}}]{gout21}%
  \BibitemOpen
  \bibfield  {author} {\bibinfo {author} {\bibfnamefont {Q.}~\bibnamefont
  {Goutaland}}, \bibinfo {author} {\bibfnamefont {F.}~\bibnamefont {van
  Wijland}}, \bibinfo {author} {\bibfnamefont {J.-B.}\ \bibnamefont
  {Fournier}}, \ and\ \bibinfo {author} {\bibfnamefont {H.}~\bibnamefont
  {Noguchi}},\ }\href {\doibase 10.1039/d1sm00027f} {\bibfield  {journal}
  {\bibinfo  {journal} {Soft Matter}\ }\textbf {\bibinfo {volume} {17}},\
  \bibinfo {pages} {5560} (\bibinfo {year} {2021})}\BibitemShut {NoStop}%
\bibitem [{\citenamefont {Noguchi}(2021{\natexlab{b}})}]{nogu21b}%
  \BibitemOpen
  \bibfield  {author} {\bibinfo {author} {\bibfnamefont {H.}~\bibnamefont
  {Noguchi}},\ }\href {\doibase 10.1039/d1sm01360b} {\bibfield  {journal}
  {\bibinfo  {journal} {Soft Matter}\ }\textbf {\bibinfo {volume} {17}},\
  \bibinfo {pages} {10469} (\bibinfo {year} {2021}{\natexlab{b}})}\BibitemShut
  {NoStop}%
\bibitem [{\citenamefont {Gov}(2018)}]{gov18}%
  \BibitemOpen
  \bibfield  {author} {\bibinfo {author} {\bibfnamefont {N.~S.}\ \bibnamefont
  {Gov}},\ }\href@noop {} {\bibfield  {journal} {\bibinfo  {journal} {Phil.
  Trans. R. Soc. B}\ }\textbf {\bibinfo {volume} {373}},\ \bibinfo {pages}
  {20170115} (\bibinfo {year} {2018})}\BibitemShut {NoStop}%
\bibitem [{\citenamefont {Wu}\ \emph {et~al.}(2018)\citenamefont {Wu},
  \citenamefont {Su}, \citenamefont {Tong}, \citenamefont {Wu},\ and\
  \citenamefont {Liu}}]{wu18}%
  \BibitemOpen
  \bibfield  {author} {\bibinfo {author} {\bibfnamefont {Z.}~\bibnamefont
  {Wu}}, \bibinfo {author} {\bibfnamefont {M.}~\bibnamefont {Su}}, \bibinfo
  {author} {\bibfnamefont {C.}~\bibnamefont {Tong}}, \bibinfo {author}
  {\bibfnamefont {M.}~\bibnamefont {Wu}}, \ and\ \bibinfo {author}
  {\bibfnamefont {J.}~\bibnamefont {Liu}},\ }\href@noop {} {\bibfield
  {journal} {\bibinfo  {journal} {Nat. Commun.}\ }\textbf {\bibinfo {volume}
  {9}},\ \bibinfo {pages} {136} (\bibinfo {year} {2018})}\BibitemShut {NoStop}%
\bibitem [{\citenamefont {Tamemoto}\ and\ \citenamefont
  {Noguchi}(2020)}]{tame20}%
  \BibitemOpen
  \bibfield  {author} {\bibinfo {author} {\bibfnamefont {N.}~\bibnamefont
  {Tamemoto}}\ and\ \bibinfo {author} {\bibfnamefont {H.}~\bibnamefont
  {Noguchi}},\ }\href {\doibase 10.1038/s41598-020-76695-x} {\bibfield
  {journal} {\bibinfo  {journal} {Sci. Rep.}\ }\textbf {\bibinfo {volume}
  {10}},\ \bibinfo {pages} {19582} (\bibinfo {year} {2020})}\BibitemShut
  {NoStop}%
\bibitem [{\citenamefont {Tamemoto}\ and\ \citenamefont
  {Noguchi}(2021)}]{tame21}%
  \BibitemOpen
  \bibfield  {author} {\bibinfo {author} {\bibfnamefont {N.}~\bibnamefont
  {Tamemoto}}\ and\ \bibinfo {author} {\bibfnamefont {H.}~\bibnamefont
  {Noguchi}},\ }\href {\doibase 10.1039/d1sm00540e} {\bibfield  {journal}
  {\bibinfo  {journal} {Soft Matter}\ }\textbf {\bibinfo {volume} {17}},\
  \bibinfo {pages} {6589} (\bibinfo {year} {2021})}\BibitemShut {NoStop}%
\bibitem [{\citenamefont {Fournier}(1996)}]{four96}%
  \BibitemOpen
  \bibfield  {author} {\bibinfo {author} {\bibfnamefont {J.-B.}\ \bibnamefont
  {Fournier}},\ }\href@noop {} {\bibfield  {journal} {\bibinfo  {journal}
  {Phys. Rev. Lett.}\ }\textbf {\bibinfo {volume} {76}},\ \bibinfo {pages}
  {4436} (\bibinfo {year} {1996})}\BibitemShut {NoStop}%
\bibitem [{\citenamefont {Kabaso}\ \emph {et~al.}(2011)\citenamefont {Kabaso},
  \citenamefont {Gongadze}, \citenamefont {Elter}, \citenamefont {van Rienen},
  \citenamefont {Gimsa}, \citenamefont {Kralj-Igli{\v{c}}},\ and\ \citenamefont
  {Igli{\v{c}}}}]{kaba11}%
  \BibitemOpen
  \bibfield  {author} {\bibinfo {author} {\bibfnamefont {D.}~\bibnamefont
  {Kabaso}}, \bibinfo {author} {\bibfnamefont {E.}~\bibnamefont {Gongadze}},
  \bibinfo {author} {\bibfnamefont {P.}~\bibnamefont {Elter}}, \bibinfo
  {author} {\bibfnamefont {U.}~\bibnamefont {van Rienen}}, \bibinfo {author}
  {\bibfnamefont {J.}~\bibnamefont {Gimsa}}, \bibinfo {author} {\bibfnamefont
  {V.}~\bibnamefont {Kralj-Igli{\v{c}}}}, \ and\ \bibinfo {author}
  {\bibfnamefont {A.}~\bibnamefont {Igli{\v{c}}}},\ }\href@noop {} {\bibfield
  {journal} {\bibinfo  {journal} {Mini Rev. Med. Chem.}\ }\textbf {\bibinfo
  {volume} {11}},\ \bibinfo {pages} {272} (\bibinfo {year} {2011})}\BibitemShut
  {NoStop}%
\bibitem [{\citenamefont {Dommersnes}\ and\ \citenamefont
  {Fournier}(1999)}]{domm99}%
  \BibitemOpen
  \bibfield  {author} {\bibinfo {author} {\bibfnamefont {P.~G.}\ \bibnamefont
  {Dommersnes}}\ and\ \bibinfo {author} {\bibfnamefont {J.-B.}\ \bibnamefont
  {Fournier}},\ }\href@noop {} {\bibfield  {journal} {\bibinfo  {journal} {Eur.
  Phys. J. B}\ }\textbf {\bibinfo {volume} {12}},\ \bibinfo {pages} {9}
  (\bibinfo {year} {1999})}\BibitemShut {NoStop}%
\bibitem [{\citenamefont {Dommersnes}\ and\ \citenamefont
  {Fournier}(2002)}]{domm02}%
  \BibitemOpen
  \bibfield  {author} {\bibinfo {author} {\bibfnamefont {P.~G.}\ \bibnamefont
  {Dommersnes}}\ and\ \bibinfo {author} {\bibfnamefont {J.-B.}\ \bibnamefont
  {Fournier}},\ }\href@noop {} {\bibfield  {journal} {\bibinfo  {journal}
  {Biophys. J.}\ }\textbf {\bibinfo {volume} {83}},\ \bibinfo {pages} {2898}
  (\bibinfo {year} {2002})}\BibitemShut {NoStop}%
\bibitem [{\citenamefont {Schweitzer}\ and\ \citenamefont
  {Kozlov}(2015)}]{schw15}%
  \BibitemOpen
  \bibfield  {author} {\bibinfo {author} {\bibfnamefont {Y.}~\bibnamefont
  {Schweitzer}}\ and\ \bibinfo {author} {\bibfnamefont {M.~M.}\ \bibnamefont
  {Kozlov}},\ }\href@noop {} {\bibfield  {journal} {\bibinfo  {journal} {PLoS
  Comput. Biol.}\ }\textbf {\bibinfo {volume} {11}},\ \bibinfo {pages}
  {e1004054} (\bibinfo {year} {2015})}\BibitemShut {NoStop}%
\bibitem [{\citenamefont {Noguchi}\ and\ \citenamefont
  {Fournier}(2017)}]{nogu17}%
  \BibitemOpen
  \bibfield  {author} {\bibinfo {author} {\bibfnamefont {H.}~\bibnamefont
  {Noguchi}}\ and\ \bibinfo {author} {\bibfnamefont {J.-B.}\ \bibnamefont
  {Fournier}},\ }\href {\doibase 10.1039/c7sm00305f} {\bibfield  {journal}
  {\bibinfo  {journal} {Soft\ Matter}\ }\textbf {\bibinfo {volume} {13}},\
  \bibinfo {pages} {4099} (\bibinfo {year} {2017})}\BibitemShut {NoStop}%
\bibitem [{\citenamefont {Noguchi}(2015)}]{nogu15b}%
  \BibitemOpen
  \bibfield  {author} {\bibinfo {author} {\bibfnamefont {H.}~\bibnamefont
  {Noguchi}},\ }\href {\doibase 10.1063/1.4931896} {\bibfield  {journal}
  {\bibinfo  {journal} {J. Chem. Phys.}\ }\textbf {\bibinfo {volume} {143}},\
  \bibinfo {pages} {243109} (\bibinfo {year} {2015})}\BibitemShut {NoStop}%
\bibitem [{\citenamefont {Tozzi}\ \emph {et~al.}(2021)\citenamefont {Tozzi},
  \citenamefont {Walani}, \citenamefont {Roux}, \citenamefont {Roca-Cusachs},\
  and\ \citenamefont {Arroyo}}]{tozz21}%
  \BibitemOpen
  \bibfield  {author} {\bibinfo {author} {\bibfnamefont {C.}~\bibnamefont
  {Tozzi}}, \bibinfo {author} {\bibfnamefont {N.}~\bibnamefont {Walani}},
  \bibinfo {author} {\bibfnamefont {A.-L.~L.}\ \bibnamefont {Roux}}, \bibinfo
  {author} {\bibfnamefont {P.}~\bibnamefont {Roca-Cusachs}}, \ and\ \bibinfo
  {author} {\bibfnamefont {M.}~\bibnamefont {Arroyo}},\ }\href {\doibase
  10.1039/d0sm01733g} {\bibfield  {journal} {\bibinfo  {journal} {Soft Matter}\
  }\textbf {\bibinfo {volume} {17}},\ \bibinfo {pages} {3367} (\bibinfo {year}
  {2021})}\BibitemShut {NoStop}%
\bibitem [{\citenamefont {Roux}\ \emph {et~al.}(2021)\citenamefont {Roux},
  \citenamefont {Tozzi}, \citenamefont {Walani}, \citenamefont {Quiroga},
  \citenamefont {Zalvidea}, \citenamefont {Trepat}, \citenamefont {Staykova},
  \citenamefont {Arroyo},\ and\ \citenamefont {Roca-Cusachs}}]{roux21}%
  \BibitemOpen
  \bibfield  {author} {\bibinfo {author} {\bibfnamefont {A.-L.~L.}\
  \bibnamefont {Roux}}, \bibinfo {author} {\bibfnamefont {C.}~\bibnamefont
  {Tozzi}}, \bibinfo {author} {\bibfnamefont {N.}~\bibnamefont {Walani}},
  \bibinfo {author} {\bibfnamefont {X.}~\bibnamefont {Quiroga}}, \bibinfo
  {author} {\bibfnamefont {D.}~\bibnamefont {Zalvidea}}, \bibinfo {author}
  {\bibfnamefont {X.}~\bibnamefont {Trepat}}, \bibinfo {author} {\bibfnamefont
  {M.}~\bibnamefont {Staykova}}, \bibinfo {author} {\bibfnamefont
  {M.}~\bibnamefont {Arroyo}}, \ and\ \bibinfo {author} {\bibfnamefont
  {P.}~\bibnamefont {Roca-Cusachs}},\ }\href {\doibase
  10.1038/s41467-021-26591-3} {\bibfield  {journal} {\bibinfo  {journal} {Nat.
  Commun.}\ }\textbf {\bibinfo {volume} {12}},\ \bibinfo {pages} {6550}
  (\bibinfo {year} {2021})}\BibitemShut {NoStop}%
\bibitem [{\citenamefont {Nascimento}\ \emph {et~al.}(2017)\citenamefont
  {Nascimento}, \citenamefont {Palffy-Muhoray}, \citenamefont {Taylor},
  \citenamefont {Virga},\ and\ \citenamefont {Zheng}}]{nasc17}%
  \BibitemOpen
  \bibfield  {author} {\bibinfo {author} {\bibfnamefont {E.~S.}\ \bibnamefont
  {Nascimento}}, \bibinfo {author} {\bibfnamefont {P.}~\bibnamefont
  {Palffy-Muhoray}}, \bibinfo {author} {\bibfnamefont {J.~M.}\ \bibnamefont
  {Taylor}}, \bibinfo {author} {\bibfnamefont {E.~G.}\ \bibnamefont {Virga}}, \
  and\ \bibinfo {author} {\bibfnamefont {X.}~\bibnamefont {Zheng}},\ }\href
  {\doibase 10.1103/PhysRevE.96.022704} {\bibfield  {journal} {\bibinfo
  {journal} {Phys. Rev. E}\ }\textbf {\bibinfo {volume} {96}},\ \bibinfo
  {pages} {022704} (\bibinfo {year} {2017})}\BibitemShut {NoStop}%
\bibitem [{\citenamefont {M{\"u}ller}\ \emph {et~al.}(2006)\citenamefont
  {M{\"u}ller}, \citenamefont {Katsov},\ and\ \citenamefont {Schick}}]{muel06}%
  \BibitemOpen
  \bibfield  {author} {\bibinfo {author} {\bibfnamefont {M.}~\bibnamefont
  {M{\"u}ller}}, \bibinfo {author} {\bibfnamefont {K.}~\bibnamefont {Katsov}},
  \ and\ \bibinfo {author} {\bibfnamefont {M.}~\bibnamefont {Schick}},\ }\href
  {\doibase 10.1016/j.physrep.2006.08.003} {\bibfield  {journal} {\bibinfo
  {journal} {Phys.\ Rep.}\ }\textbf {\bibinfo {volume} {434}},\ \bibinfo
  {pages} {113} (\bibinfo {year} {2006})}\BibitemShut {NoStop}%
\bibitem [{\citenamefont {Venturoli}\ \emph {et~al.}(2006)\citenamefont
  {Venturoli}, \citenamefont {Sperotto}, \citenamefont {Kranenburg},\ and\
  \citenamefont {Smit}}]{vent06}%
  \BibitemOpen
  \bibfield  {author} {\bibinfo {author} {\bibfnamefont {M.}~\bibnamefont
  {Venturoli}}, \bibinfo {author} {\bibfnamefont {M.~M.}\ \bibnamefont
  {Sperotto}}, \bibinfo {author} {\bibfnamefont {M.}~\bibnamefont
  {Kranenburg}}, \ and\ \bibinfo {author} {\bibfnamefont {B.}~\bibnamefont
  {Smit}},\ }\href {\doibase 10.1016/j.physrep.2006.07.006} {\bibfield
  {journal} {\bibinfo  {journal} {Phys.\ Rep.}\ }\textbf {\bibinfo {volume}
  {437}},\ \bibinfo {pages} {1} (\bibinfo {year} {2006})}\BibitemShut {NoStop}%
\bibitem [{\citenamefont {Noguchi}(2009)}]{nogu09}%
  \BibitemOpen
  \bibfield  {author} {\bibinfo {author} {\bibfnamefont {H.}~\bibnamefont
  {Noguchi}},\ }\href {\doibase 10.1143/JPSJ.78.041007} {\bibfield  {journal}
  {\bibinfo  {journal} {J.\ Phys.\ Soc.\ Jpn.}\ }\textbf {\bibinfo {volume}
  {78}},\ \bibinfo {pages} {041007} (\bibinfo {year} {2009})}\BibitemShut
  {NoStop}%
\bibitem [{\citenamefont {Arkhipov}\ \emph {et~al.}(2008)\citenamefont
  {Arkhipov}, \citenamefont {Yin},\ and\ \citenamefont {Schulten}}]{arkh08}%
  \BibitemOpen
  \bibfield  {author} {\bibinfo {author} {\bibfnamefont {A.}~\bibnamefont
  {Arkhipov}}, \bibinfo {author} {\bibfnamefont {Y.}~\bibnamefont {Yin}}, \
  and\ \bibinfo {author} {\bibfnamefont {K.}~\bibnamefont {Schulten}},\
  }\href@noop {} {\bibfield  {journal} {\bibinfo  {journal} {Biophys.\ J.}\
  }\textbf {\bibinfo {volume} {95}},\ \bibinfo {pages} {2806} (\bibinfo {year}
  {2008})}\BibitemShut {NoStop}%
\bibitem [{\citenamefont {Yu}\ and\ \citenamefont {Schulten}(2013)}]{yu13}%
  \BibitemOpen
  \bibfield  {author} {\bibinfo {author} {\bibfnamefont {H.}~\bibnamefont
  {Yu}}\ and\ \bibinfo {author} {\bibfnamefont {K.}~\bibnamefont {Schulten}},\
  }\href@noop {} {\bibfield  {journal} {\bibinfo  {journal} {PLoS Comput.
  Biol.}\ }\textbf {\bibinfo {volume} {9}},\ \bibinfo {pages} {e1002892}
  (\bibinfo {year} {2013})}\BibitemShut {NoStop}%
\bibitem [{\citenamefont {Simunovic}\ \emph {et~al.}(2013)\citenamefont
  {Simunovic}, \citenamefont {Srivastava},\ and\ \citenamefont
  {Voth}}]{simu13}%
  \BibitemOpen
  \bibfield  {author} {\bibinfo {author} {\bibfnamefont {M.}~\bibnamefont
  {Simunovic}}, \bibinfo {author} {\bibfnamefont {A.}~\bibnamefont
  {Srivastava}}, \ and\ \bibinfo {author} {\bibfnamefont {G.~A.}\ \bibnamefont
  {Voth}},\ }\href@noop {} {\bibfield  {journal} {\bibinfo  {journal} {Proc.\
  Natl.\ Acad.\ Sci.\ USA}\ }\textbf {\bibinfo {volume} {110}},\ \bibinfo
  {pages} {20396} (\bibinfo {year} {2013})}\BibitemShut {NoStop}%
\bibitem [{\citenamefont {G{\'o}mez-Llobregat}\ \emph
  {et~al.}(2016)\citenamefont {G{\'o}mez-Llobregat}, \citenamefont
  {El{\'i}as-Wolff},\ and\ \citenamefont {Lind{\'e}n}}]{gome16}%
  \BibitemOpen
  \bibfield  {author} {\bibinfo {author} {\bibfnamefont {J.}~\bibnamefont
  {G{\'o}mez-Llobregat}}, \bibinfo {author} {\bibfnamefont {F.}~\bibnamefont
  {El{\'i}as-Wolff}}, \ and\ \bibinfo {author} {\bibfnamefont {M.}~\bibnamefont
  {Lind{\'e}n}},\ }\href@noop {} {\bibfield  {journal} {\bibinfo  {journal}
  {Biophys. J.}\ }\textbf {\bibinfo {volume} {110}},\ \bibinfo {pages} {197}
  (\bibinfo {year} {2016})}\BibitemShut {NoStop}%
\bibitem [{\citenamefont {Olinger}\ \emph {et~al.}(2016)\citenamefont
  {Olinger}, \citenamefont {Spangler}, \citenamefont {Kumar},\ and\
  \citenamefont {Laradji}}]{olin16}%
  \BibitemOpen
  \bibfield  {author} {\bibinfo {author} {\bibfnamefont {A.~D.}\ \bibnamefont
  {Olinger}}, \bibinfo {author} {\bibfnamefont {E.~J.}\ \bibnamefont
  {Spangler}}, \bibinfo {author} {\bibfnamefont {P.~B.~S.}\ \bibnamefont
  {Kumar}}, \ and\ \bibinfo {author} {\bibfnamefont {M.}~\bibnamefont
  {Laradji}},\ }\href@noop {} {\bibfield  {journal} {\bibinfo  {journal}
  {Faraday Discuss.}\ }\textbf {\bibinfo {volume} {186}},\ \bibinfo {pages}
  {265} (\bibinfo {year} {2016})}\BibitemShut {NoStop}%
\bibitem [{\citenamefont {Takemura}\ \emph {et~al.}(2017)\citenamefont
  {Takemura}, \citenamefont {Hanawa-Suetsugu}, \citenamefont {Suetsugu},\ and\
  \citenamefont {Kitao}}]{take17}%
  \BibitemOpen
  \bibfield  {author} {\bibinfo {author} {\bibfnamefont {K.}~\bibnamefont
  {Takemura}}, \bibinfo {author} {\bibfnamefont {K.}~\bibnamefont
  {Hanawa-Suetsugu}}, \bibinfo {author} {\bibfnamefont {S.}~\bibnamefont
  {Suetsugu}}, \ and\ \bibinfo {author} {\bibfnamefont {A.}~\bibnamefont
  {Kitao}},\ }\href@noop {} {\bibfield  {journal} {\bibinfo  {journal} {Sci.
  Rep.}\ }\textbf {\bibinfo {volume} {7}},\ \bibinfo {pages} {6808} (\bibinfo
  {year} {2017})}\BibitemShut {NoStop}%
\bibitem [{\citenamefont {Mahmood}\ \emph {et~al.}(2019)\citenamefont
  {Mahmood}, \citenamefont {Noguchi},\ and\ \citenamefont {Okazaki}}]{mahm19}%
  \BibitemOpen
  \bibfield  {author} {\bibinfo {author} {\bibfnamefont {M.~I.}\ \bibnamefont
  {Mahmood}}, \bibinfo {author} {\bibfnamefont {H.}~\bibnamefont {Noguchi}}, \
  and\ \bibinfo {author} {\bibfnamefont {K.}~\bibnamefont {Okazaki}},\ }\href
  {\doibase 10.1038/s41598-019-51202-z} {\bibfield  {journal} {\bibinfo
  {journal} {Sci. Rep.}\ }\textbf {\bibinfo {volume} {9}},\ \bibinfo {pages}
  {14557} (\bibinfo {year} {2019})}\BibitemShut {NoStop}%
\bibitem [{\citenamefont {Ramakrishnan}\ \emph {et~al.}(2013)\citenamefont
  {Ramakrishnan}, \citenamefont {{Sunil Kumar}},\ and\ \citenamefont
  {Ipsen}}]{rama13}%
  \BibitemOpen
  \bibfield  {author} {\bibinfo {author} {\bibfnamefont {N.}~\bibnamefont
  {Ramakrishnan}}, \bibinfo {author} {\bibfnamefont {P.~B.}\ \bibnamefont
  {{Sunil Kumar}}}, \ and\ \bibinfo {author} {\bibfnamefont {J.~H.}\
  \bibnamefont {Ipsen}},\ }\href@noop {} {\bibfield  {journal} {\bibinfo
  {journal} {Biophys. J.}\ }\textbf {\bibinfo {volume} {104}},\ \bibinfo
  {pages} {1018} (\bibinfo {year} {2013})}\BibitemShut {NoStop}%
\bibitem [{\citenamefont {Behera}\ \emph {et~al.}(2021)\citenamefont {Behera},
  \citenamefont {Kumar}, \citenamefont {Akram},\ and\ \citenamefont
  {Sain}}]{behe21}%
  \BibitemOpen
  \bibfield  {author} {\bibinfo {author} {\bibfnamefont {A.}~\bibnamefont
  {Behera}}, \bibinfo {author} {\bibfnamefont {G.}~\bibnamefont {Kumar}},
  \bibinfo {author} {\bibfnamefont {S.~A.}\ \bibnamefont {Akram}}, \ and\
  \bibinfo {author} {\bibfnamefont {A.}~\bibnamefont {Sain}},\ }\href@noop {}
  {\bibfield  {journal} {\bibinfo  {journal} {Soft Matter}\ }\textbf {\bibinfo
  {volume} {17}},\ \bibinfo {pages} {7953} (\bibinfo {year}
  {2021})}\BibitemShut {NoStop}%
\bibitem [{\citenamefont {Noguchi}(2014)}]{nogu14}%
  \BibitemOpen
  \bibfield  {author} {\bibinfo {author} {\bibfnamefont {H.}~\bibnamefont
  {Noguchi}},\ }\href {\doibase 10.1209/0295-5075/108/48001} {\bibfield
  {journal} {\bibinfo  {journal} {EPL}\ }\textbf {\bibinfo {volume} {108}},\
  \bibinfo {pages} {48001} (\bibinfo {year} {2014})}\BibitemShut {NoStop}%
\bibitem [{\citenamefont {Noguchi}(2016{\natexlab{a}})}]{nogu16}%
  \BibitemOpen
  \bibfield  {author} {\bibinfo {author} {\bibfnamefont {H.}~\bibnamefont
  {Noguchi}},\ }\href {\doibase 10.1038/srep20935} {\bibfield  {journal}
  {\bibinfo  {journal} {Sci.\ Rep.}\ }\textbf {\bibinfo {volume} {6}},\
  \bibinfo {pages} {20935} (\bibinfo {year} {2016}{\natexlab{a}})}\BibitemShut
  {NoStop}%
\bibitem [{\citenamefont {Noguchi}(2016{\natexlab{b}})}]{nogu16a}%
  \BibitemOpen
  \bibfield  {author} {\bibinfo {author} {\bibfnamefont {H.}~\bibnamefont
  {Noguchi}},\ }\href {\doibase 10.1103/PhysRevE.93.052404} {\bibfield
  {journal} {\bibinfo  {journal} {Phys.\ Rev.\ E}\ }\textbf {\bibinfo {volume}
  {93}},\ \bibinfo {pages} {052404} (\bibinfo {year}
  {2016}{\natexlab{b}})}\BibitemShut {NoStop}%
\bibitem [{\citenamefont {Noguchi}(2017)}]{nogu17a}%
  \BibitemOpen
  \bibfield  {author} {\bibinfo {author} {\bibfnamefont {H.}~\bibnamefont
  {Noguchi}},\ }\href {\doibase 10.1039/c7sm01375b} {\bibfield  {journal}
  {\bibinfo  {journal} {Soft\ Matter}\ }\textbf {\bibinfo {volume} {13}},\
  \bibinfo {pages} {7771} (\bibinfo {year} {2017})}\BibitemShut {NoStop}%
\bibitem [{\citenamefont {Noguchi}(2019{\natexlab{a}})}]{nogu19a}%
  \BibitemOpen
  \bibfield  {author} {\bibinfo {author} {\bibfnamefont {H.}~\bibnamefont
  {Noguchi}},\ }\href {\doibase 10.1038/s41598-019-48102-7} {\bibfield
  {journal} {\bibinfo  {journal} {Sci.\ Rep.}\ }\textbf {\bibinfo {volume}
  {9}},\ \bibinfo {pages} {11721} (\bibinfo {year}
  {2019}{\natexlab{a}})}\BibitemShut {NoStop}%
\bibitem [{\citenamefont {Smith}\ \emph {et~al.}(2004)\citenamefont {Smith},
  \citenamefont {Sackmann},\ and\ \citenamefont {Seifert}}]{smit04}%
  \BibitemOpen
  \bibfield  {author} {\bibinfo {author} {\bibfnamefont {A.-S.}\ \bibnamefont
  {Smith}}, \bibinfo {author} {\bibfnamefont {E.}~\bibnamefont {Sackmann}}, \
  and\ \bibinfo {author} {\bibfnamefont {U.}~\bibnamefont {Seifert}},\ }\href
  {\doibase 10.1103/PhysRevLett.92.208101} {\bibfield  {journal} {\bibinfo
  {journal} {Phys. Rev. Lett.}\ }\textbf {\bibinfo {volume} {92}},\ \bibinfo
  {pages} {208101} (\bibinfo {year} {2004})}\BibitemShut {NoStop}%
\bibitem [{\citenamefont {Tanemura}\ and\ \citenamefont
  {Matsumoto}(1997)}]{tane97}%
  \BibitemOpen
  \bibfield  {author} {\bibinfo {author} {\bibfnamefont {M.}~\bibnamefont
  {Tanemura}}\ and\ \bibinfo {author} {\bibfnamefont {T.}~\bibnamefont
  {Matsumoto}},\ }\href@noop {} {\bibfield  {journal} {\bibinfo  {journal}
  {Zeit. Krist.}\ }\textbf {\bibinfo {volume} {212}},\ \bibinfo {pages} {637}
  (\bibinfo {year} {1997})}\BibitemShut {NoStop}%
\bibitem [{\citenamefont {Seifert}(1997)}]{seif97}%
  \BibitemOpen
  \bibfield  {author} {\bibinfo {author} {\bibfnamefont {U.}~\bibnamefont
  {Seifert}},\ }\href@noop {} {\bibfield  {journal} {\bibinfo  {journal} {Adv.\
  Phys.}\ }\textbf {\bibinfo {volume} {46}},\ \bibinfo {pages} {13} (\bibinfo
  {year} {1997})}\BibitemShut {NoStop}%
\bibitem [{\citenamefont {Svetina}(2009)}]{svet09}%
  \BibitemOpen
  \bibfield  {author} {\bibinfo {author} {\bibfnamefont {S.}~\bibnamefont
  {Svetina}},\ }\href@noop {} {\bibfield  {journal} {\bibinfo  {journal}
  {ChemPhysChem}\ }\textbf {\bibinfo {volume} {10}},\ \bibinfo {pages} {2769}
  (\bibinfo {year} {2009})}\BibitemShut {NoStop}%
\bibitem [{\citenamefont {Shiba}\ and\ \citenamefont {Noguchi}(2011)}]{shib11}%
  \BibitemOpen
  \bibfield  {author} {\bibinfo {author} {\bibfnamefont {H.}~\bibnamefont
  {Shiba}}\ and\ \bibinfo {author} {\bibfnamefont {H.}~\bibnamefont
  {Noguchi}},\ }\href {\doibase 10.1103/PhysRevE.84.031926} {\bibfield
  {journal} {\bibinfo  {journal} {Phys. Rev. E}\ }\textbf {\bibinfo {volume}
  {84}},\ \bibinfo {pages} {031926} (\bibinfo {year} {2011})}\BibitemShut
  {NoStop}%
\bibitem [{\citenamefont {Noguchi}(2019{\natexlab{b}})}]{nogu19}%
  \BibitemOpen
  \bibfield  {author} {\bibinfo {author} {\bibfnamefont {H.}~\bibnamefont
  {Noguchi}},\ }\href {\doibase 10.1063/1.5113646} {\bibfield  {journal}
  {\bibinfo  {journal} {J.\ Chem.\ Phys.}\ }\textbf {\bibinfo {volume} {151}},\
  \bibinfo {pages} {094903} (\bibinfo {year} {2019}{\natexlab{b}})}\BibitemShut
  {NoStop}%
\bibitem [{\citenamefont {Noguchi}(2011)}]{nogu11}%
  \BibitemOpen
  \bibfield  {author} {\bibinfo {author} {\bibfnamefont {H.}~\bibnamefont
  {Noguchi}},\ }\href {\doibase 10.1063/1.3541246} {\bibfield  {journal}
  {\bibinfo  {journal} {J.\ Chem.\ Phys.}\ }\textbf {\bibinfo {volume} {134}},\
  \bibinfo {pages} {055101} (\bibinfo {year} {2011})}\BibitemShut {NoStop}%
\bibitem [{\citenamefont {Hukushima}\ and\ \citenamefont
  {Nemoto}(1996)}]{huku96}%
  \BibitemOpen
  \bibfield  {author} {\bibinfo {author} {\bibfnamefont {K.}~\bibnamefont
  {Hukushima}}\ and\ \bibinfo {author} {\bibfnamefont {K.}~\bibnamefont
  {Nemoto}},\ }\href@noop {} {\bibfield  {journal} {\bibinfo  {journal} {J.\
  Phys.\ Soc.\ Jpn.}\ }\textbf {\bibinfo {volume} {65}},\ \bibinfo {pages}
  {1604} (\bibinfo {year} {1996})}\BibitemShut {NoStop}%
\bibitem [{\citenamefont {Okamoto}(2004)}]{okam04}%
  \BibitemOpen
  \bibfield  {author} {\bibinfo {author} {\bibfnamefont {Y.}~\bibnamefont
  {Okamoto}},\ }\href@noop {} {\bibfield  {journal} {\bibinfo  {journal} {J.
  Mol. Graph. Model.}\ }\textbf {\bibinfo {volume} {22}},\ \bibinfo {pages}
  {425} (\bibinfo {year} {2004})}\BibitemShut {NoStop}%
\end{thebibliography}
\end{document}